\documentclass[12pt,a4paper,notitlepage]{article}
\usepackage[utf8]{inputenc}
\usepackage[english]{babel}
\usepackage[T1]{fontenc}
\usepackage{hyperref}
\usepackage{amsmath}
\usepackage{stmaryrd}
\usepackage{amsfonts}
\usepackage{amssymb}
\usepackage{color} 
\usepackage{caption}
\usepackage{amsmath}
\usepackage{relsize,exscale}
\usepackage{array,multirow,makecell}

\usepackage[toc,page]{appendix}

\setcellgapes{1pt}
\makegapedcells
\newcolumntype{R}[1]{>{\raggedleft\arraybackslash }b{#1}}
\newcolumntype{L}[1]{>{\raggedright\arraybackslash }b{#1}}
\newcolumntype{C}[1]{>{\centering\arraybackslash }b{#1}}

\newtheorem{theorem}{Theorem}
\newtheorem{definition}{Definition}
\newtheorem{proposition}{Proposition}

\newtheorem{claim}{Claim}
\newtheorem{corollary}{Corollary}

\newcommand{\sym}{\mathrm{Sym}}

\newtheorem{lemma}{Lemma}
\usepackage{enumerate}
\usepackage{subfig}
\usepackage{graphicx}

\newcommand{\cL}{{\mathcal L}}

\newcommand{\beq}{\begin{equation}}
\newcommand{\eeq}{\end{equation}}
\newcommand{\bea}{\begin{eqnarray}}
\newcommand{\eea}{\end{eqnarray}}
\definecolor{mygray}{gray}{0.3}

\newcommand{\bes}{\begin{eqnarray}}
\newcommand{\ees}{\end{eqnarray}}

\newcommand\restr[2]{{
  \left.\kern-\nulldelimiterspace 
  #1 
  \vphantom{\big|} 
  \right|_{#2} 
  }}

\usepackage{multicol}
\usepackage{amssymb}

\makeatletter
\begin{document}
\begin{center}
\textbf{\Large{Large-$d$ behavior of the Feynman amplitudes for a just-renormalizable tensorial group field theory}}\\
\vspace{15pt}

{\large Vincent Lahoche$^a$\footnote{vincent.lahoche@cea.fr}  \,\,and 
Dine Ousmane Samary$^{a,b}$\footnote{dine.ousmanesamary@cipma.uac.bj}
 }
\vspace{15pt}

a)\,  Commissariat à l'\'Energie Atomique (CEA, LIST),
 8 Avenue de la Vauve, 91120 Palaiseau, France

b)\, Facult\'e des Sciences et Techniques (ICMPA-UNESCO Chair),
Universit\'e d'Abomey-
Calavi, 072 BP 50, Benin
\vspace{0.5cm}
\end{center}
\begin{center}
\begin{abstract}
This paper aims at giving a novel approach to investigate the behavior of the renormalization group flow for tensorial group field theories to all orders of the perturbation theory. From an appropriate choice of the kinetic kernel, we build an infinite family of just-renormalizable models, for tensor fields with arbitrary rank $d$. Investigating the large $ d$ limit, we show that the self-energy melonic amplitude is decomposed as a product of loop-vertex functions, depending only on dimensionless mass. The corresponding melonic amplitudes may be mapped as trees in the so-called Hubbard-Stratonivich representation, and we show that only trees with edges of different colors survive in the large $d$-limit. These two key features allow us to resum the perturbative expansion for self-energy, providing an explicit expression for arbitrary external momenta in terms of Lambert function. Finally, inserting this resumed solution into the Callan-Symanzik equations, and taking into account the strong relation between two and four-point functions arising from melonic Ward-Takahashi identities, we then deduce an explicit expression for relevant and marginal $\beta$-functions, valid to all orders of the perturbative expansion. By investigating the solutions of the resulting flow, we conclude about the non-existence of any fixed point in the investigated region of the full phase space.
\end{abstract}
\end{center}

\newpage
\tableofcontents
\pagebreak
\section{Introduction}

Tensorial group field theories (TGFT) were born a decade, after the merger between group field theories (GFT) and colored tensor models, both appeared in the quantum gravity context as peculiar field theoretical frameworks \cite{Oriti:2005jr}-\cite{Hawking:1979pi}.

GFTs on one hand, arise from loop quantum gravity (LQG) and spin-foam theory, as a promising way to generate spin-foam amplitudes as Feynman amplitude for a field theory defined over a group manifold, with specific non-local interactions. GFTs are introduced by the so-called Boulatov model for three-dimensional gravity and arise as a way to implement simplicial decomposition for a pseudo-manifold, including discrete connection through a specific invariance called closure constraint \cite{Ooguri:1991ib}-\cite{Ooguri:1992tw}. More recently, it was shown that GFTs may be viewed in a complementary way as a second quantized version of spin network states of loop quantum gravity \cite{Hawking:1978jz}-\cite{Perez:2000ec}, quantum excitations being interpreted as spin-network nodes, which can be combined to build LQG states. The fact that GFT provides a field theoretical framework for spin foam theory, allowing to use of standard field theoretical methods, represents great progress in itself, explaining the success of the approach in the last decade. Among these successes highlighting the power of the field theoretical framework, the most important one at this day is undoubtedly the results obtained in the context of the quantum cosmology \cite{Oriti:2018qty}-\cite{Gielen:2017eco}, as a mathematical incarnation of the \textit{geometrogenesis} scenario, the space-time Universe is viewed as a condensate of quantum gravity building blocks. Since the first approaches described a homogeneous Universe and recovered the classical Friedman equations in the classical limit; These recent results showed that inhomogeneous effects can be described as well in the same condensation scenario, considering a multi-condensate state, and the resulting evolution equations for perturbations are in strong agreement with the classical results, so far from the Planck scale. There is no doubt that these attractive results are the beginning of a long history of quantum cosmology, where GFTs will be demonstrated their power. \\

Although they have been originally introduced in the GFT context to cancel some pathologies as singular topologies' proliferation, colored tensor models (TM) on the other hand may be viewed as the generalization of the random matrix which is the discrete approach to random geometry for two-dimensional manifolds \cite{Brezin:1992yc}-\cite{DiFrancesco:1993cyw}. The breakthrough of colored TM, and probably the reason for their success, is certainly the existence of a tractable power counting, allowing to build a $1/N$ expansion like for matrix models \cite{Gurau:2011aq}-\cite{Gurau:2011xq}. The role played by the genus in the case of matrix theory is now replaced by a new quantity called the Gurau degree. Despite the fact that the \textit{Gurau degree} is not a topological invariant, in contrast to the genus, it provides a good definition of leading order graphs, which corresponds to a vanish Gurau degree and leads to the so-called melonic diagrams. In contrast to planar graphs, the melons obey a recursive definition allowing them to map as $d$-ary trees and then become easy to count. Beyond the vectors and matrices, tensors, and particularly the melonic diagrams may provide for future developments, many applications are far from quantum gravity. \\

TGFTs are built by merging some aspect of these two approaches (GFTs and TMs) such that: The fields remain defined on a group manifold, but the interactions inherit their structure from tensor models. A new notion of locality usually referred to as \textit{traciality} replaces the simplicial constraint from which GFT interactions are historically constructed. As for tensor models, this locality principle allows defining a power counting, and then addressing the question of renormalization see \cite{Samary:2012bw}-\cite{Lahoche:2015ola} and references therein. In standard quantum and statistical field theory context, the canonical notions of scale and locality arise from the background space-time itself. For GFT however, there is no background to support these notions, and the scales, which are required for standard renormalization procedure have to be defined extrinsically as well. This is given in practice through a modification of the kinetic action \cite{Geloun:2011cy}, the notion of scale arising from the spectrum of the kinetic kernel, usually a linear combination of the identity and the Laplace-Beltrami operator, both defined over the group manifold. Rigorous BPHZ theorems have been proved for such a kind of field theories, from which potentially interesting theories have been classified from their perturbative just-renormalizability \cite{Samary:2012bw}-\cite{Lahoche:2015ola}. Finally, nonperturbative renomalization group aspects have been addressed for these models through the popular Wetterich-Morris formalism, which is the most suitable to deal with the specific locality of TGFTs in order to investigate the strong coupling regime \cite{Carrozza:2014rba}-\cite{Lahoche:2019orv}. Some non-Gaussian fixed points, reminiscent of phase transitions, have been obtained for all the investigated models; which have been pointed out to be in strong agreement with the phase condensation at the heart of the geometrogenesis scenario \cite{Mandrysz:2018sle}-\cite{Markopoulou:2007jf}. More recently, a series of papers \cite{Lahoche:2018vun}-\cite{Lahoche:2019orv} took into account Ward-Takahashi identities in the renormalization group equations. Indeed, for the models without closure constraint, the strong violations of Ward identities for the discovered fixed points particularly for marginal quartic interactions have been checked at the level of just-renormalizable's interactions \cite{Lahoche:2018ggd}. \\

\noindent
In this paper, we address the question of the existence or not of such a fixed point from a completely different point of view, through an exploration of the large rank limit of a just-renormalizable family of models. We show that in this limit, only a subfamily of melons survives, providing a good recurrence relation for Feynman amplitudes. Taking into account only the $1PI$ two-point Feynman amplitudes, this recurrence relation can be solved in terms of Lambert functions, leading to an explicit expression for the two-point function to all orders of the perturbative expansion. From this explicit solution, and taking into account the strong relation between two and four-point functions arising from Ward identities in the melonic approximation, we solve the Callan-Symanzik (C-S) equation, and deduce an explicit expression for relevant and marginal $\beta$-functions. Finally, investigating these solutions, we show that no-fixed point occurs in the considered region of the full space of couplings. \\

\noindent
Note that the resumed two-point function that we obtain in our computation provides a solution, in a suitable limit, to the so-called closed melonic equation, firstly introduced in \cite{Samary:2014tja}-\cite{Sanchez:2017gxt} by direct inspirations of the Grosse and Wulkenhaar works for non-commutative field theory \cite{Grosse:2019qps}-\cite{Grosse:2012uv}. Up to the leading order melonic diagrams, these equations are reputed to be very hard to solve. In the first paper on this subject \cite{Samary:2014tja}-\cite{Sanchez:2017gxt}, the authors only addressed a perturbative solution of a just-renormalizable model, up to order six. The same equation has been considered for a tensorial group field theory endowed with a specific gauge invariance called closure constraint \cite{Lahoche:2015ola} on which only the perturbative solution is also given. In the same reference paper, and from a BPHZ theorem, the authors argued in favour of the existence of a solution to all orders of the perturbative expansion. Recently, strong progress has been achieved in \cite{Pascalie:2019yxd}, where an explicit solution has been found for a divergent free model. However, to this day, no such solution exists for just-renormalizable models. In this paper, we show that in a suitable large $d$ limit, an explicit analytic solution can be found for the melonic closed equation. \\

\noindent
In detail, the outline of this paper is the following. In section \ref{section2} we define the model and introduce the useful materials used in the rest of the paper, from which more information may be found in the Appendix and the list of references cited above. Among the key results of this section, we get a strong relation between four and two-point melonic functions, arising from Ward identity, which can be translated as a local relation between $\beta$-function along the RG flow. In section \ref{section3} we investigate the large rank behavior of the Feynman amplitude; first, we provide the one and two loops computation in order to get the recurrence relation which could help to a generalization to arbitrary $n$-loops. Then, using the recurrence relation on the perturbative expansion, we derive the same result at all orders. Explicitly, we show that Feynman amplitudes for two-point graphs may be factorized as a product of functions, whose, one depends on the external momentum. In the melonic sector, and for large $d$, each of these amplitudes can be indexed by a planar rooted tree with edges of different colors rather than one obtained in the ordinary Feynman graph. Then, by summing over all such trees, we get an explicit expression for self-energy, from which we can deduce the $\ beta$ functions.

\section{Preliminaries }\label{section2}

In this section, we introduce the notations and the formalism that we will use for the rest of this paper and recall some important definitions and results (additional details could be found in standard references \cite{Lahoche:2018oeo}). A second time, we build explicitly a just-renormalizable family of models for arbitrary rank $d$, some complementary results about the proof of renormalizability could be found in Appendix \ref{App1}. Finally, we discuss the existence of a non-trivial relation between four and two-point functions, holding to all orders in the perturbative expansion, and show explicitly that the information of the renormalization group flow in the deep ultraviolet version reduces to the information of the self-energy at zero momenta and its first derivative.

\subsection{Just-renormalizable Abelian TGFT in rank $d$}\label{section21}

TGFT that we consider in this paper describes two fields $\varphi$ and $\bar\varphi$, both defined on $d$ copies of a compact Lie group manifold $G$: $\varphi,\bar{\varphi}: (G)^d\to \mathbb{C}$. For our purpose we focus on the Abelian manifolds, choosing $G=U(1)$ and the fields are then defined on the $d$-dimensional torus. From the trivial exponential map $ \theta \mapsto e^{i\theta}\in U(1)$, rather than functions of group elements, the fields can be understood as functions of the angle variables $\theta\in[\,0,2\pi[$, and we denote as $\varphi(\theta_1,\cdots,\theta_d)\equiv \varphi(\vec{\theta})$ the field arguments (same for the field $\bar{\varphi}$). Moreover, instead of focusing on the group (or Lie-Algebra) representation, it is more convenient to use the Fourier representation, the Fourier components $T$ and $\bar T$ of $\varphi$ and $\bar\varphi$ respectively, being formal \textit{tensors} of rank $d$ i.e. discrete maps from $\mathbb{Z}^d$ to $\mathbb{C}$. We denote their components as $T_{\vec p}$ and $\bar{T}_{\vec{p}}$, with $\vec p=(p_1,\cdots,p_d)\in\mathbb{Z}^d$. At the classical level, tensors are described by the classical action $S[T,\bar{T}]$, which is assumed to be quartic for our purpose:
\begin{equation}
S[T,\bar{T}]=\sum_{\vec{p}} \bar{T}_{\vec{p}} \, \mathcal{K}(\vec{p}\,) \, T_{\vec{p}} +\lambda \sum_{i} \sum_{\vec{p}_1,\cdots,\vec{p}_4}\mathcal{V}^{(i)}_{\vec{p}_1,\vec{p}_2,\vec{p}_3,\vec{p}_4}T_{\vec{p}_1}\bar{T}_{\vec{p}_2}T_{\vec{p}_3}\bar{T}_{\vec{p}_4}\,,\label{classicaction}
\end{equation}
where $\lambda$ denotes the coupling constant and $\mathcal{V}^{(i)}_{\vec{p}_1,\vec{p}_2,\vec{p}_3,\vec{p}_4}$ the \textit{vertex tensor}, i.e. a product of Kronecker deltas which dictates how the tensor indices are contracted together. This coupling tensor is obviously not unique, we distinguish the different choices of them by the subscript $i$. Note that with our definition of the classical action, all these components are chosen with the same coupling. The only constraint over $\mathcal{V}^{(i)}_{\vec{p}_1,\vec{p}_2,\vec{p}_3,\vec{p}_4}$ comes from the tensoriality criterion, ensuring that any index of a field $T$ has to be contracted with an index of a field $\bar{T}$. In this paper, we focus on the quartic melonic model, for which the set of coupling tensors write explicitly as:
\begin{equation}
\mathcal{V}^{(i)}_{\vec{p}_1,\vec{p}_2,\vec{p}_3,\vec{p}_4}:=\delta_{p_{1i}p_{4i}}\delta_{p_{2i}p_{3i}}\prod_{j\neq i} \delta_{p_{1j}p_{2j}}\delta_{p_{3j}p_{4j}}\,.\label{melobubble}
\end{equation}
All the interactions whose tensor couplings decompose in this way i.e. who do not factorize as a product over subsets of indices for some $T$ and $\bar{T}$, is called \textit{bubbles}; and the couplings defined from \eqref{melobubble} are known as quartic melonic bubbles. Bubbles may be fruitfully pictured as bipartite regular-colored graphs, a representation that we will use abundantly in the rest of this paper. The rule to build the correspondence is the following. To each $T$ and $\bar{T}$ fields, we associate respectively black and white nodes; each of them being hooked to $d$ colored half edges. These $d$ edges, corresponding to the $d$ components of the tensor are then joined following the path provided by the interaction tensor, any half edge of color $c$ starting from a black node being hooked to a half edge of the same color hooked to a white node. For melonic quartic couplings, we have:
\begin{equation}
\sum_{\vec{p}_1,\cdots,\vec{p}_4}\mathcal{V}^{(i)}_{\vec{p}_1,\vec{p}_2,\vec{p}_3,\vec{p}_4}T_{\vec{p}_1}\bar{T}_{\vec{p}_2}T_{\vec{p}_3}\bar{T}_{\vec{p}_4} \equiv \vcenter{\hbox{\includegraphics[scale=0.6]{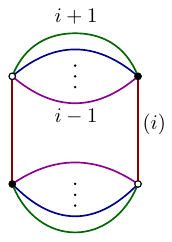} }}\,.
\end{equation}
The interacting part of the classical action being fixed, let us define the kinetic action. Renormalization requires that the kinetic kernel $\mathcal{K}(\vec{p}\,)$ must have a non-trivial spectrum in order to provide a canonical notion of scale. The standard choice, motivated by radiative computation in GFTs \cite{Geloun:2011cy}, involves the Laplace-Beltrami operator over the group manifold $G^d$. For $G=U(1)$, this Laplace-Beltrami operator is diagonal in the Fourier representation. Setting the kinetic kernel to be:
\begin{equation}
\mathcal{K}(\vec{p}\,)= \vec{p}\,^{2}+m^{2}\,,\label{standardkin}
\end{equation}
and the corresponding field theory, with quartic-melonic interactions, have been stated to be just-renormalizable for $d=5$ \cite{Samary:2012bw}. In this paper, we will relax the power of the Laplacian term, and consider the slight generalization:
\begin{equation}
\mathcal{K}(\vec{p}\,):= \vec{p}\,^{2\eta}+m^{2\eta}\,,
\end{equation}
where $\eta$ is chosen to be a \textit{positive half-integer}, $\eta=n/2$; $n\in \mathbb{N}$, and the notation $ \vec{p}\,^{2\eta}$ simply means $ \vec{p}\,^{2\eta}:= \sum_{i=1}^d \vert p_i\vert^{2\eta}$. The motivation for such a deformation with respect to the standard choice \eqref{standardkin} arises from the power counting which will be discussed in detail below. The choice of $\eta$ influence strongly the divergent degree, and may be fixed such that the model remains just-renormalizable for arbitrary rank $d$. Note that we introduced a $\eta$-dependent power on the mass term, in the hope to get the same \textit{canonical dimension} for a just-renormalizable theory. Finally let us remark that we do not consider the closure constraint in our models, even if it is considered a crucial ingredient for GFTs. \\

\begin{figure}
\begin{center}
\includegraphics[scale=1]{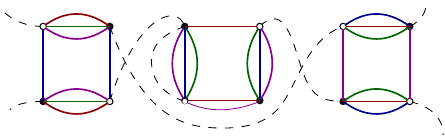}
\captionof{figure}{A typical Feynman graph with three vertices and four external lines. The propagator lines hooked to black and white nodes are dotted lines, and open dotted lines are external lines. }\label{fig1}
\end{center}
\end{figure}

\noindent
The statistical theory, introducing integration over ‘‘thermal'' fluctuations is defined from the classical action \eqref{classicaction} by the path integral:
\begin{equation}
\mathcal{Z}(\lambda)=\int d\mu_C[T,\bar{T}] e^{-S_{\text{int}}[T,\bar{T}]}\,,
\end{equation}
where $S_{\text{int}}$ designates the quartic part of the classical action, and $d\mu_C$ is the normalized Gaussian measure for the propagator $C$, defined as:
\begin{equation}
\int d\mu_C[T,\bar{T}]T_{\vec{p}}\bar{T}_{\vec{p}\,^\prime} =\frac{\Theta(\Lambda^{2\eta}-\vec{p}\,^{2\eta}\,)}{\vec{p}\,^{2\eta}+m^{2\eta}} \delta_{\vec{p}\,\vec{p}\,^\prime}\,.
\end{equation}
where we introduced the step Heaviside function $\Theta$ to prevent UV divergences. Up to the standard permutation of sums and integrals (which in general is not well defined), the perturbative expansion in powers of the coupling $\lambda$ organizes as a sum of amplitudes indexed of Feynman graphs, such that the
1PI-connected $N$-point function $\mathcal{S}_N$ (which depends on $N$ external momenta) writes as:
\begin{equation}
\mathcal{S}_N=\sum_{\mathcal{G}_N} \frac{(-\lambda)^{V(\mathcal{G}_N)}}{s(\mathcal{G}_N)}\mathcal{A}_{\mathcal{G}_N}\,,
\end{equation}
where the sum run over the connected graphs with $N$ external edges, $V(\mathcal{G}_N)$ designates the number of melonic vertices of the graph $\mathcal{G}_N$, and $s(\mathcal{G}_N)$ is a combinatorial factor coming from the Wick theorem\footnote{Note that it does not reduce to the dimension of the automorphism group of the considered graph, due to the absence of the factor $1/4$ in front of $\lambda$ in the classical action.}. Due to the specific combinatorial structure of the interactions, these Feynman graphs $\mathcal{G}_N$ are $2$ simplex rather than ordinary graphs, i.e. have sets of vertices, edges and faces. Such a typical graph with four external edges is given in Figure \ref{fig1} on which the Wick contractions are pictured as dotted edges between black and white nodes, to whose we attribute the color $0$, so that Feynman graphs become $d+1$ bipartite graphs. We recall that faces are bicolored cycles, including necessarily the color $0$, and may be open or closed, respectively for external and internal faces.

\begin{figure}[!h]
\includegraphics[scale=0.8]{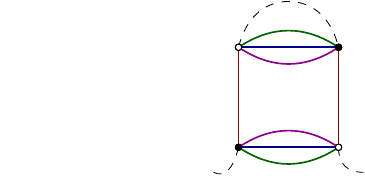} \hfill
\includegraphics[scale=0.8]{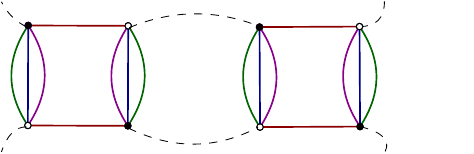}
\caption{Leading order contributions for 1PI $2$ and $4$ point functions. The figures have been drawn for $d=4$.}\label{fig2}
\end{figure}

Renormalizability of the quartic melonic models has been studied extensively, especially for $\eta=1$ and $\eta=1/2$, i.e. for linear and quadratic kinetic kernels \cite{Samary:2012bw},\cite{BenGeloun:2012pu}, \cite{Geloun:2013saa}. In particular, just-renormalizability has been proved for $\eta=1$ and $d=5$. Here we just fix the parameter $\eta$ differently, such that the model becomes just-renormalizable for arbitrary rank $d$. Such a fixation requires the knowledge of power counting which takes place through the important and useful techniques called multi-scale analysis and may help to establish the solid power-counting theorem. Note that in practice, the presence of the parameter $\eta$ does not change significantly the main steps of the proofs given in the previous references. Some details was reproduced in appendix \ref{App1}, and the result is the following:
\begin{proposition}
The power counting $\omega(\mathcal G)$ for a Feynman graph $\mathcal{G}$ with $L(\mathcal{G})$ internal lines and $F(\mathcal{G})$ internal faces is given by:
\begin{equation}
\omega(\mathcal G)=-2\eta L(\mathcal G)+F(\mathcal G)\,.
\end{equation}
\end{proposition}
At this step, we choose $\eta$ such that the leading order graphs, i.e. the melonic graphs appear in the renormalization procedure and are just the relevant graphs for the computation of the beta functions in the UV sector. In appendix \ref{App1}, we prove that a sufficient condition is: $d>3$, ensuring that for any deviation from the melonic sector, the deleted faces never compensate the variation over the number of internal (dotted) edges. Due to the recursive structure of the melonic graphs, as well recalled in appendix \ref{App1}, we can prove the following statement, which links together the number of internal dotted edges, vertices, and internal faces:
\begin{equation}\label{F}
F(\mathcal G)=(d-1)(L(\mathcal G)-V(\mathcal G)+1)\,.
\end{equation}
Combining this expression with the following topological relation arising from the valence of the quartic vertices:
\begin{equation}
4V(\mathcal G)=2L(\mathcal G)+N(\mathcal G)\,,
\end{equation}
where $N(\mathcal G)$ designates the number of external lines, we deduce that melonic diagrams diverge as:
\begin{equation}\label{dimrel}
\omega(G)=\big[(d-1)-4\eta\big]V+\bigg[(d-1)-\bigg(\frac{d-1}{2}-\eta\bigg)N\bigg]\,.
\end{equation}
For a just-renormalizable theory, the divergent degrees must have to be independent of the vertex number \footnote{If the divergent degree decrease with the number of vertices, the situation is still interesting, because it means that there are only a finite set of divergent graphs, which could be subtracted to rend the theory finitely. This corresponds to a super-renormalizable theory. On the other hand, the divergent degree increases with the number of vertices, and an infinite number of counter-terms is required to make the theory well-defined in the UV. Fixing an infinite number of counter-terms, or equivalently an infinite number of "initial conditions" break the predictivity of the theory, which is said to be non-renormalizable.} so that UV-divergences can be removed from a finite set of counter-terms, even if the number of graphs is infinite. This condition fixes the value of $\eta$ as:
\begin{equation}
\eta=\frac{d-1}{4}\,.
\end{equation}
For the standard field theories defined on space-time, the just-renormalizability property is closely related to the dimension of the coupling, which has to vanish for just-renormalizable theories. For TGFTs, there is no meaning to talk about dimension, because there is no space-time background, and the sums over $\mathbb{Z}^d$ are dimensionless. An intrinsic notion of dimension however emerges from the renormalization group flow itself, following the behavior of the renormalization group trajectories. In the vicinity of the Gaussian point, the \textit{canonical dimension} is then fixed from the behavior of the leading order Feynman amplitudes -- the dimension being fixed from the scaling of the leading order quantum corrections with respect to some UV cut-off. This notion is of great interest for the rest of this paper and we provide here a brief explanation of its origin. As an illustration, let us consider the first quantum corrections for the mass parameter, which provides from the diagram pictured in Figure \ref{fig2} (on left) below. If we denote by $L_1$ the loop involved in the diagram, the mass correction takes the form:
\begin{equation}
\delta m^{2\eta}=\lambda K_1L_1 \,,
\end{equation}
where $K_1$ is a numerical (cut-off independent) factor. Denoting by $[x]$ the dimension of the quantity $x$, we get the first relation:
\begin{equation}
[m^{2\eta}]=[\lambda]+[L_1]\,.
\end{equation}
A second relation comes from the first radiative correction for the $4$-points function, see Figure \ref{fig2} (on right). denote by $L_2$ the loop of length $2$ involved on the diagram, we get the relation : $[\lambda]=2[\lambda]+[L_2]$. Now, observe that $L_1$ and $L_2$ have the same number of internal faces, i.e. $d-1$. Their respective scaling then becomes:
\begin{equation}
\omega(L_1)=-2\eta+(d-1)\,,\qquad \omega(L_2)=-4\eta+(d-1)\,,
\end{equation}
as a result:
\begin{equation}
[m^{2\eta}]=[\lambda]+(d-1)-2\eta\,,\quad [\lambda]=2[\lambda]+(d-1)-4\eta \,,
\end{equation}
leading to:
\begin{equation}
[\lambda]=4\eta-(d-1)\,,\quad [m^{2\eta}]=2\eta\,.
\end{equation}
Note that the dimension of $m^{2\eta}$ is fixed to be $2\eta$, as suggested by the notations. Moreover, if the theory is renormalizable, $[\lambda]=0$, as expected from standard quantum field theory.

Then we come to the following definition which ends this section
\begin{definition} \textbf{Boundary and heart vertices and faces}\\\label{def1}

\noindent
$\bullet$ Any vertex hooked with an external edge is said to be a boundary vertex. Other vertices are called heart vertices\\

\noindent
$\bullet$ Any external faces running through a single external vertex are said to be boundary external faces. \\

\noindent
$\bullet$ Any external face running through at least one heart vertex is said to be a heart external face.
\end{definition}

\subsection{Exact relation between effective (melonic) vertex and wave function}\label{section23}

Because of their recursive definition, there exist strong relations between melonic diagrams with two, four or an arbitrary number of external edges, such that the melonic sector is entirely determined by the knowledge of the melonic self-energy. This section aims to establish the exact relation holding between two and four-point functions, and the corresponding relations between counter-terms.\\

\noindent
The melonic self-energy $\Sigma(\vec{p}\,)$, i.e. whose perturbative expansion keeps only the melonic diagrams, is related to the two-point function $\Gamma^{(2)}(\vec{p}\,)$ as in ordinary field theory:
\begin{equation}
\Gamma^{(2)}(\vec{p}\,)=\vec{p}\,^{2\eta}+m^{2\eta}-\Sigma(\vec{p}\,)\,.
\end{equation}
Moreover, as a direct consequence of the recursive definition of the melonic diagrams, the melonic self-energy obeys a closed equation, which as we announced in the introduction is reputed to be very difficult to solve. The proof of these closed equations and the main corollary statements can be found in \cite{Lahoche:2015ola},\cite{Samary:2014oya}-\cite{Samary:2014tja}. To summarize:
\begin{proposition}\textbf{Closed equation for self-energy.}\label{closedequation}
Let $\Sigma(\vec{p}\,)$ be the melonic self-energy, whose Feynman expansion involves only melonic diagrams. Then, all the variables are completely decoupled and $\Sigma(\vec{p}\,)$ is a sum of $d$ independent terms, one per variables:
\begin{equation}
\Sigma(\vec{p}\,)=:\sum_{i=1}^d \tau(p_i)\,,
\end{equation}
where the function $\tau:\mathbb{Z}\to\mathbb{R}$ has a single argument and satisfies the closed equation:
\begin{equation}
\tau(p):=-2\lambda\sum_{\vec{q}}\delta_{pq_1}\frac{\Theta(\Lambda^{2\eta}-\vec{q}\,^{2\eta}\,)}{\vec{q}\,^{2\eta}+m^{2\eta}-\sum_{i=1}^d\tau(q_i)}\,.\label{closed}
\end{equation}
\end{proposition}
\noindent
Note that $\tau(p)$ only depends on $p^2$. From the power counting theorem, only the two and four-point melonic diagrams diverge and then require renormalization. Moreover, in the deep UV limit, the knowledge of the counter-terms allows computing the beta functions. As we will see, the unitary symmetry of the action, explicitly broken by the kinetic kernel implies the existence of a strong relation between four and two-point functions through the standard Ward-Takahashi identity. More precisely, we have the following statement:

\begin{proposition} \textbf{Zero-momenta Ward identity.}\label{Wardid}
Let $\gamma^{(4)}:=\Gamma^{(4)}_{\vec{0},\vec{0},\vec{0},\vec{0}}$ be the zero momenta 1PI melonic $4$-point function. In the continuum limit, $\gamma^{(4)}$ is related to the first derivative of the $2$-point melonic function $\Gamma^{(2)}(\vec{p}\,)$ as:
\begin{equation}
\frac{1}{2}\gamma^{(4)}\mathcal{L}(1+\partial\tau) =\tau^\prime(0)\,,\label{Ward}
\end{equation}
with the notation $\tau^\prime(0):=\partial\tau/\partial p_1^{2\eta}\vert_{p_1=0}$, and the loop $\mathcal{L}$ as the ‘‘boundary contribution" $\partial \tau$ are defined as:
\begin{equation}
\mathcal{L}:=\sum_{\vec{q}}\delta_{q_10}[\Gamma^{(2)}(\vec{q}\,)]^{-2}\,,
\end{equation}
\begin{equation}
\mathcal{L}\partial\tau:= \sum_{\vec{q}} \delta_{q_10}(\vec{q}\,^{2\eta}+m^{2\eta})[\Gamma^{(2)}(\vec{q}\,)]^{-2}\delta(\Lambda^{2\eta}-\vec{q}\,^{2\eta}\,)\,.
\end{equation}
\end{proposition}

\noindent
\textit{Proof.} Let us consider the unitary transformations $\textbf{U}\in\mathcal{U}^{\times d}$ acting independently over each components of the tensors $T$ and $\bar{T}$. $\textbf{U}$ is a $d$-dimensional vector $\textbf{U}=(U_1,U_2,\cdots,U_d)$ whose components $U_i$ are unitary matrices acting on the indices of color $i$. The action of $\textbf{U}$ on the two tensors are defined as (we sum over repeated indices):
\begin{align}
\textbf{U}[T]_{p_1,p_2,\cdots,p_d}&:=[U_1]_{p_1q_1}[U_2]_{p_2q_2}\cdots[U_d]_{p_dq_d}T_{q_1,q_2,\cdots,q_d}\\
\textbf{U}[\bar{T}]_{p_1,p_2,\cdots,p_d}&:=[U_1^*]_{p_1q_1}[U_2^*]_{p_2q_2}\cdots[U_d^*]_{p_dq_d}\bar{T}_{q_1,q_2,\cdots,q_d}\,,
\end{align}
where $*$ means complex conjugation. Obviously, $\sum_{\vec{p}}\bar{T}_{\vec{p}}T_{\vec{p}}$ and any higher valence tensorial interactions are invariant under any such transformations. Then:
\begin{equation}
\textbf{U}[S_{\text{int}}]=S_{\text{int}}\,.
\end{equation}
However, this is not the case for the kinetic term, due to the non-trivial propagator, which explicitly breaks the unitary invariance. Now, let us consider the two point function $\langle\bar{T}_{\vec{p}}T_{\vec{q}}\rangle$. It is tempting to think that it transforms like a representation of $\textbf{U}\otimes \textbf{U}^*$. Indeed, even if the kinetic term does not transform like a tensorial invariant, the integral:
\begin{equation}
\langle\bar{T}_{\vec{p}}T_{\vec{q}}\rangle:=\int d\mu_C \bar{T}_{\vec{p}}\,T_{\vec{q}}\,e^{-S_{int}[\bar{T},T]}\,,
\end{equation}
do not depend on the broken symmetry transformation of the kinetic term because of the formal translation invariance of the Lebesgue integration measure, and in fact, it has to be invariant under any unitary transformation. Furthermore, $\langle\bar{T}_{\vec{p}}\,T_{\vec{q}}\rangle$ transforms like a trivial representation of $\mathcal{U}^{\times d}\otimes \mathcal{U}^{*\times d}$. This can be translated in an infinitesimal point of view considering an infinitesimal transformation $U=\mathbb{I}+i\epsilon$, where $\epsilon=\epsilon^\dagger$ is a hermitian operator and $\mathbb{I}$ the identity operator. At the first order in $\epsilon$, we get:
\begin{equation}
\textbf{U}=\textbf{I}+\sum_i\vec{\epsilon}_i\,,
\end{equation}
where $\textbf{I}:=\mathbb{I}^{\otimes d}$ and $\vec{\epsilon}_i=\mathbb{I}^{\otimes (i-1)}\otimes\epsilon_i\otimes\mathbb{I}^{d-i+1}$. Then, the invariance of $\langle\bar{T}_{\vec{p}}T_{\vec{q}}\rangle$ simply means that $\vec{\epsilon}_i[\langle\bar{T}_{\vec{p}}T_{\vec{q}}\rangle]=0$. Expanding this relation at the leading order in $\epsilon_i$, and due to the symmetry $\vec{\epsilon}_i[S_{int}]=0$, we get:
\begin{equation}
\int \vec{\epsilon}_i[d\mu_C] \bar{T}_{\vec{p}}T_{\vec{q}}\,e^{-S_{int}[\bar{T},T]}+\int d\mu_C \vec{\epsilon}_i[\bar{T}_{\vec{p}}T_{\vec{q}}\,]e^{-S_{int}[\bar{T},T]}=0\,.
\end{equation}
Each term can be computed separately. The variation of the covariance requires to be carefully derived because the propagator $C$ is not invertible on $\mathbb{Z}^d$ due to the $\Theta$-function $\Theta(\Lambda^{2\eta}-\vec{p}\,^{2\eta}\,)$. The computation of the variation then requires regularization of the infinite coming from $1/\Theta$. The variation of the second term however can be computed straightforwardly. From:
\begin{align}
\nonumber\vec{\epsilon}_i[\bar{T}_{\vec{p}}T_{\vec{q}}]&=-\epsilon_{p_ip_i^\prime}^*\bar{T}_{\vec{p}\,^\prime}\prod_{j\neq i} \delta_{p_jp_j^\prime}T_{\vec{q}}+\bar{T}_{\vec{p}}\, \epsilon_{q_iq_i^\prime}T_{\vec{q}\,^\prime}\prod_{j\neq i} \delta_{q_jq_j^\prime}\\\nonumber
&= \, \epsilon_{q_iq_i^\prime}\bar{T}_{\vec{p}}\prod_{j\neq i} \delta_{q_jq_j^\prime}T_{\vec{q}\,^\prime}-\epsilon_{p_i^\prime p_i}\bar{T}_{\vec{p}\,^\prime}\prod_{j\neq i} \delta_{p_jp_j^\prime}T_{\vec{q}}\\
&=\bar{T}_{\vec{p}}T_{\vec{q}_{\bot_i}\cup \{q_i^\prime\}} \epsilon_{q_iq_i^\prime}-\bar{T}_{\vec{p}_{\bot_i}\cup \{p_i^\prime\}}T_{\vec{q}}\,\epsilon_{p_i^\prime p_i}\,,
\end{align}
where $\vec{p}_{\bot_i}:=\vec{p}\,/\{p_i\}\in\mathbb{Z}^{d-1}$. Integrating with the measure $d\mu_C e^{-S_{int}[\bar{T},T]}$, and after restricting our computation on the perturbative sector, it is obvious that $\langle \bar{T}_{\vec{p}}T_{\vec{q}}\rangle\propto \delta_{\vec{p}\vec{q}}$ due to the momentum conservation along all the external faces. Then, setting:
\begin{equation}
\langle \bar{T}_{\vec{p}}T_{\vec{q}}\rangle = G(\vec{p}\,)\delta_{\vec{p}\vec{q}}\,,
\end{equation}
we get:
\begin{equation}
\int d\mu_C \vec{\epsilon}_i[\bar{T}_{\vec{p}}T_{\vec{q}}\,]e^{-S_{int}[\bar{T},T]}=\delta_{\vec{p}_{\bot_i}\vec{q}_{\bot_i}}[G(\vec{p}\,)-G(\vec{q}\,)]\epsilon_{q_ip_i}\,.\label{var1}
\end{equation}
Now let us focus on the variation of the measure $d\mu_C$. As explained before, we have to regularise the $\ Theta$ function occurring on the propagator. We use the well know relation between Heaviside and Dirac function: $\theta^\prime=\delta$, and the Gaussian representation of the finite range $\delta$-function:
\begin{equation}
\delta_a(x):=\frac{1}{a\sqrt{\pi}}e^{-x^2/a^2}\,,
\end{equation}
which goes to the standard Dirac function\footnote{More precisely, it goes to the delta distribution on the space of test functions $\mathcal{D}(\mathbb{R})$.} when $a\to 0$. Then we get the following limit:
\begin{equation}
\theta_a(x):=\frac{1}{a\sqrt{\pi}}\int_{-\infty}^x \,e^{-y^2/a^2}dy \,,\quad \lim\limits_{a \to 0} \theta_a=\Theta\,.
\end{equation}
Therefore, defining $C_0^{-1}(\vec{p}\,):=\vec{p}\,^2+m^2$, our regularized propagator can be written $C_a(\vec{p}\,)=\theta_a(\Lambda^2-\vec{p}\,^2\,)C_0(\vec{p}\,)$ such that the Gaussian measure and its variation are written as:
\begin{equation}
d\mu_{C_a}:=e^{-\sum_{\vec{p}}\bar{T}_{\vec{p}}\,C^{-1}_a(\vec{p}\,)T_{\vec{p}}}\,,\quad \vec{\epsilon}_i[d\mu_{C_a}]=-\vec{\epsilon}_i\Big[\sum_{\vec{p}}\bar{T}_{\vec{p}}\,C^{-1}_a(\vec{p}\,)T_{\vec{p}}\,\Big]d\mu_{C_a}\,.
\end{equation}
The variation of the kinetic term then becomes:
\begin{equation}
\vec{\epsilon}_i\Big[\sum_{\vec{p}}\bar{T}_{\vec{p}}\,C^{-1}_a(\vec{p}\,)T_{\vec{p}}\,\Big]=\sum_{\vec{p},\vec{q}}\epsilon_{q_ip_i}\delta_{\vec{p}_{\bot_i}\vec{q}_{\bot_i}}[C_a^{-1}(\vec{p}\,)-C_a^{-1}(\vec{q}\,)]\bar{T}_{\vec{p}}T_{\vec{q}}\,.\label{var2}
\end{equation}
By considering the results \eqref{var1} and \eqref{var2}, we get:
\begin{align*}
\sum_{\vec{r},\vec{s}}\epsilon_{r_is_i}\delta_{\vec{r}_{\bot_i}\vec{s}_{\bot_i}}&[C_a^{-1}(\vec{r}\,)-C_a^{-1}(\vec{s}\,)]\langle\bar{T}_{\vec{r}}T_{\vec{s}}\bar{T}_{\vec{p}}T_{\vec{q}}\rangle\\
&\qquad\qquad=\sum_{r_i,s_i}\left[\delta_{\vec{p}_{\bot_i}\vec{q}_{\bot_i}}(G(\vec{p}\,)-G(\vec{q}\,))\delta_{r_iq_i}\delta_{s_ip_i}\right]\epsilon_{r_is_i}\,,
\end{align*}
or, because of the arbitrariness of the infinitesimal transformation $\epsilon$:
\begin{equation}\label{satsat}
\sum_{\vec{r}_{\bot_i},\vec{s}_{\bot_i}}\delta_{\vec{r}_{\bot_i}\vec{s}_{\bot_i}}[C_a^{-1}(\vec{r}\,)-C_a^{-1}(\vec{s}\,)]\langle\bar{T}_{\vec{r}}T_{\vec{s}}\bar{T}_{\vec{p}}T_{\vec{q}}\rangle=\left[\delta_{\vec{p}_{\bot_i}\vec{q}_{\bot_i}}(G(\vec{p}\,)-G(\vec{q}\,))\delta_{r_iq_i}\delta_{s_ip_i}\right]\,.
\end{equation}
Let $\Gamma_{\vec{p}_1,\vec{p}_2,\vec{p}_3,\vec{p}_4}^{(4)}$ be the 1PI four points function defined by the following relation
\begin{equation}
\langle\bar{T}_{\vec{r}}T_{\vec{s}}\bar{T}_{\vec{p}}T_{\vec{q}}\rangle=:\left(-\Gamma_{\vec{r},\vec{s},\vec{p},\vec{q}}^{(4)}\,G(\vec{p}\,)G(\vec{q}\,)+\delta_{\vec{r}\vec{p}}\,\delta_{\vec{s}\vec{q}}\right)G(\vec{r}\,)G(\vec{s}\,)\,.
\end{equation}
Expression \eqref{satsat} becomes:
\begin{align}
\nonumber\sum_{\vec{r}_{\bot_i},\vec{s}_{\bot_i}}\delta_{\vec{r}_{\bot_i}\vec{s}_{\bot_i}}[C_a^{-1}(\vec{r}\,)-&C_a^{-1}(\vec{s}\,)]\,G(\vec{r}\,)G(\vec{s}\,)\left[-\Gamma_{\vec{r},\vec{s},\vec{p},\vec{q}}^{(4)}\,+\Gamma^{(2)}(\vec{p}\,)\Gamma^{(2)}(\vec{q}\,)\delta_{\vec{r}\vec{p}}\,\delta_{\vec{s}\vec{q}}\right]\\
&=\left[\delta_{\vec{p}_{\bot_i}\vec{q}_{\bot_i}}(\Gamma^{(2)}(\vec{q}\,)-\Gamma^{(2)}(\vec{p}\,))\delta_{r_iq_i}\delta_{s_ip_i}\right]\,,\label{wardstep}
\end{align}
with $\Gamma^{(2)}(\vec{p}\,):=1/G(\vec{p}\,)$. Using the proposition \ref{cormelons} in the Appendix \ref{App1} and taking into account the leading order contributions, $\Gamma_{\vec{r},\vec{s},\vec{p},\vec{q}}^{(4)}$ is such that any diagrams in its Feynman expansion have two heart external faces of the same color, running through the interior of the graph. As a consequence, the leading contributions for $\Gamma^{(4)}$ may be decomposed into a sum indexed by a single color like the free energy $\Sigma$ :
\begin{equation}
\Gamma^{(4)}:=\sum_{i=1}^d\Gamma^{(4)\,,i}\,,
\end{equation}
Moreover, from the same proposition \ref{cormelons}, in addition to these two heart external faces, we have $(d-1)$ boundary external faces of length $1$ per external vertices (in the case when we have only one vertex, it can be considered like an external vertex because external lines are hooked to him). Then, a moment of reflection shows that the leading order $4$-point function must have the following structure:
\begin{equation}
\Gamma^{(4)\,,i}_{\vec{p}_1,\vec{p}_2,\vec{p}_3,\vec{p}_4}=\gamma_{p_{1i}p_{3i}}^{(4)} \left(\mathcal{V}^{(i)}_{\vec{p}_1,\vec{p}_2,\vec{p}_3,\vec{p}_4}+\vec{p}_1\leftrightarrow \vec{p}_3\right)=:\gamma_{p_{1i}p_{3i}}^{(4)}\sym\mathcal{V}^{(i)}_{\vec{p}_1,\vec{p}_2,\vec{p}_3,\vec{p}_4},,
\end{equation}
where the last term comes from the Wick theorem. As a result, only the component $\Gamma^{(i)}$ contributes significantly to \eqref{wardstep} at leading order. Also, only a single term in $\sym\mathcal{V}^{(i)}$ has to be retained. Then setting $q_i=r_i$ and $p_i=s_i$ in a first time, and $\vec{p}=\vec{q}\to\vec{0}$ in a second time, \eqref{wardstep} becomes at leading order:
\begin{equation}
\frac{1}{2}\left(\sum_{\vec{r}}\delta_{r_10} \frac{dC^{-1}_a}{dr_1^{2\eta}}(\vec{r}\,)G^2(\vec{r}\,)\right)\times \gamma^{(4)}=-\frac{\partial }{\partial p_1^{2\eta}}\Gamma^{(2)}(\vec{0})+\frac{dC^{-1}_a}{dp_1^{2\eta}}(\vec{0}\,)\,.\label{wardstep2}
\end{equation}
where $\gamma^{(4)}=2\gamma^{(4)}_{00}$. From the definition:
\begin{equation}
\frac{dC^{-1}_a}{dr_1^{2\eta}}(\vec{r}\,)=\left(1+(\vec{r}\,^{2\eta}+m^{2\eta})\frac{\theta^\prime_a}{\theta_a}(\Lambda^{2\eta}-\vec{r}\,^{2\eta}\,)\right)\theta_a^{-1}(\Lambda^{2\eta}-\vec{r}\,^{2\eta}\,)\,.
\end{equation}
The derivative on the right-hand side requires the explicit expression for $\Gamma^{(2)}$. The effective propagator $G$ is obtained from $C_a$ and $\Sigma$ as a geometric progression:
\begin{equation}
G=C_a+C_a\Sigma C_a+C_a\Sigma C_a\Sigma C_a+\cdots= \frac{1}{1-C_a\Sigma}C_a \,,
\end{equation}
explicitly:
\begin{equation}
G(\vec{p}\,)=\frac{\theta_a(\Lambda^{2\eta}-\vec{p}\,^{2\eta})}{\vec{p}\,^{2\eta}+m^{2\eta}-\theta_a(\Lambda^{2\eta}-\vec{p}\,^{2\eta}\,)\Sigma(\vec{p}\,)}\,,
\end{equation}
and we get
\begin{equation}
\Gamma^{(2)}(\vec{p}\,)=\theta_a^{-1}(\Lambda^{2\eta}-\vec{p}\,^{2\eta}\,)(\vec{p}\,^{2\eta}+m^{2\eta})-\Sigma(\vec{p}\,)\,.
\end{equation}
We deduce that
\begin{equation}
\frac{\partial \Gamma^{(2)}}{\partial p_1^{2\eta}}(\vec{0}\,)-\frac{dC^{-1}_a}{dp_1^{2\eta}}(\vec{0}\,)=- \frac{\partial }{\partial p_1^{2\eta}} \Sigma(\vec{0}\,)\,.
\end{equation}
Finally, taking into account the factor $\theta_a^2$ coming from $G^2$, and by chosing $a$ to $0$, the equation \eqref{wardstep2} writes as:
\begin{equation}
\frac{1}{2}\mathcal{L}(1+\partial \tau)\times \gamma^{(4)}=\frac{\partial }{\partial p_1^{2\eta}} \Sigma(\vec{0}\,)\,.
\end{equation}
With the decomposition $\Sigma(\vec{p}\,)=\sum_i\tau(p_i)$ the proof of the proposition is therefore completed .
\begin{flushright}
$\square$
\end{flushright}

\begin{corollary}
\textbf{Exact relation between $\tau$ and $\gamma^{(4)}$. }
The zero-momenta melonic function $\gamma^{(4)}$ and the loop $\mathcal{L}$ are related as:
\begin{equation}
\gamma^{(4)}=\frac{4\lambda}{1+2\lambda \mathcal{L}}\,.
\end{equation}
\end{corollary}
\noindent
\textit{Proof.} The proof is straightforward. From the closed equation for self energy \eqref{closed}, we deduce an expression for $\tau^\prime$ involving $\mathcal{L}$:
\begin{equation}
\tau^\prime(0)=2\lambda \mathcal{L}(1-\tau^\prime(0)+\partial\tau)\to \tau^\prime(0)=\frac{2\lambda\mathcal{L}(1+\partial\tau)}{1+2\lambda\mathcal{L}}\,,
\end{equation}
Then, inserting this equation in equation \eqref{Ward} of proposition \ref{Wardid}, and after simplification of the factors $(1+\partial\tau)$, we deduce the corollary.
\begin{flushright}
$\square$
\end{flushright}

\noindent
Now let us define the functional action with the counter-terms which will be free for divergences. Denoting as $Z$, $Z_m$ and $Z_\lambda$ the counter-terms respectively for field strength, mass and coupling, such that the renormalized classical action, writing as:
\begin{equation}
S[T,\bar{T}]=\sum_{\vec{p}} \bar{T}_{\vec{p}} \,(Z\vec{p}\,^{2\eta}+Z_m m^{2\eta}) \, T_{\vec{p}} +Z_\lambda\lambda \sum_{i} \sum_{\vec{p}_1,\cdots,\vec{p}_4}\mathcal{V}^{(i)}_{\vec{p}_1,\vec{p}_2,\vec{p}_3,\vec{p}_4}T_{\vec{p}_1}\bar{T}_{\vec{p}_2}T_{\vec{p}_3}\bar{T}_{\vec{p}_4}\,,\label{classicactionren}
\end{equation}
The existence of such a set of counter-term is ensured by the renormalizability theorem. Another point of view is the behavior of effective vertex with the UV cut-off. To be more precise, $\gamma^{(4)}$ can be interpreted as an effective coupling $\lambda_{eff}$:
\begin{equation}
\lambda_{eff}=z_\lambda \lambda\,,\quad z_\lambda := \frac{1}{1+2\lambda\mathcal{L}}\,,
\end{equation}
Moreover, the relation between $\tau(0)$, $\tau^\prime(0)$ and the effective mass and wave functions can be easily found from the definition of $\Gamma^{(2)}$. We have:
\begin{equation}
\Gamma^{(2)}(\vec{p}\,)=\vec{p}\,^{2\eta}+m^{2\eta}-\Sigma(\vec{p}\,)=(1-\tau^\prime(0))\vec{p}\,^{2\eta}+(m^{2\eta}-d\times\tau(0))+\mathcal{O}(\vec{p}\,^{2\eta})\,,
\end{equation}
from which we deduce the effective wave function $Z_{eff}$ and the effective mass $m_{eff}^2$:
\begin{equation}
Z_{eff}:=1-\tau^\prime(0)=1-\frac{2\lambda\mathcal{L}(1+\partial\tau)}{1-2\lambda\mathcal{L}}=z_\lambda(1-2\lambda\mathcal{L}\partial\tau),\,\quad
m_{eff}^{2\eta}:=m^{2\eta}-d\times \tau(0)\,.
\end{equation}
\noindent
One may expect that the boundary term $\partial\tau$ introduces a spurious dependence on the UV regularization, especially in regard to the limit $a \to 0$. Indeed, the product of two distributions is not well defined in general. This is especially the case of the product $\delta\, \Theta$; and we have to be careful when we take the limit before or after the computation of the integrals. In this case, however, the limit can be well defined remembering that momenta are discrete variables and that all the derivatives are formal continuum limits of finite differences. Thus, taking the limit $a\to 0$ before the continuum limit, we have to provide a sense for integrals of the form $-\int dx \delta(1-x) G(\Theta(1-x))$; which is the limit for $\epsilon\sim 1/\Lambda^{2\eta} \to 0$ of something that:
\begin{equation}
\int dx[ \Theta(1+\epsilon-x)- \Theta(1-x)] G(\Theta(1-x))
\end{equation}
for some regular function $G(y)$. In the interval $x\in [1,1+\epsilon]$, $\Theta(1-x)$ must vanish. Therefore:
\begin{equation}
\int dx[ \Theta(1+\epsilon-x)- \Theta(1-x)] G(\Theta(1-x))\to \epsilon\, G(0) \int dx \delta(1-x)\,.
\end{equation}
Note that the result depends on the convention used to compute the derivative. Indeed, with the convention$ \Theta(1-x)- \Theta(1-\epsilon-x)$ -- which is identical for ordinary functions; we get the limit:
\begin{equation}
\int dx[ \Theta(1-x)- \Theta(1-\epsilon-x)] G(\Theta(1-x))\to \epsilon \,G(1) \int dx \delta(1-x)\,.
\end{equation}
In this paper, we use the first convention, because it offers the advantage to cancel the $\Sigma$ dependence of the denominator of the boundary term, which simplify as:
\begin{equation}
\mathcal{L}\partial\tau=\frac{1}{\Lambda\,^{2\eta}+m^{2\eta}}\sum_{\vec{q}} \delta_{q_10}\delta(\Lambda^{2\eta}-\vec{q}\,^{2\eta})\,.
\end{equation}
In the deep UV sector on which we focus in this paper, $\Lambda$ becomes large, and the continuum limit can be considered with variables $x_i=q_i/\Lambda$. We get:
\begin{equation}
\sum_{\vec{q}} \delta_{q_10}\delta(\Lambda^{2\eta}-\vec{q}\,^{2\eta})\approx 2^{d-1}\int_{\mathbb{R}^{+d-1}} d\textbf{x} \delta(1-\textbf{x}^{2\eta})\,,
\end{equation}
and from Appendix \ref{App2},
\begin{equation}
\mathcal{L}\partial\tau \approx \frac{\iota(d)}{1+\bar{m}^{2\eta}}\,, \label{conventionboundary}
\end{equation}
where we defined the dimensionless mass $\bar{m}=m/\Lambda$ such that:
\begin{equation}
Z_{eff}=z_\lambda\left(1-2\lambda \frac{\iota(d)}{1+\bar{m}^{2\eta}} \right)\,.
\end{equation}
\subsection{Melonic renormalization group equations}

The renormalization group equations (RGE) are the infinitesimal translation of a common feature of just-renormalizable theories. In the deep UV, and neglecting the contributions of inessential couplings, any change of fundamental cut-off, $\Lambda\to \Lambda^\prime$ may be exactly compensated by a change of field strength, relevant and marginal couplings -- up to corrections of order $1/\Lambda$. The infinitesimal incarnation of this feature is the so-called C-S equation, which writes as \cite{Callan:1970yg}-\cite{Symanzik:1971vw}:
\begin{equation}
\bigg(\frac{\partial}{\partial t}+\beta \frac{\partial}{\partial \lambda}+\beta_m \frac{\partial}{\partial m^{2\eta}}-\frac{N}{2}\gamma\bigg)\Gamma^{(N)}_{\vec{p}_1,\vec{p}_2,\cdots,\vec{p}_N}=0\,,\quad \forall N\,,
\end{equation}
Where $\partial/\partial t:= \Lambda \partial/\partial \Lambda$; $\beta$ and $\beta_m$ are \textit{beta functions} for quartic coupling and mass, and $\gamma$ is the \textit{anomalous dimension}. By considering the explicit expression of $\Gamma^{(2)}(\vec{0}\,)$, $\partial\Gamma^{(2)}/\partial p_1^2 (\vec{0}\,)$ and $\Gamma^{(4)}_{\vec{0},\vec{0},\vec{0},\vec{0}}\equiv\gamma^{(4)}$, and taking into account the strong relation arising from Ward identity, we deduce the statement:
\begin{proposition}
In the deep UV limit ($\Lambda\gg 1$), and with boundary term given by equation \eqref{conventionboundary}, the $\beta$-functions $\beta$, $\beta_m$ for coupling and mass; and the anomalous dimension $\gamma$ are related as:
\begin{equation}
\beta=\gamma\lambda\left(1-\frac{2\lambda\iota(d)}{1+\bar{m}^{2\eta}}\right)+\frac{2\lambda^2\iota(d)}{(1+\bar{m}^{2\eta})^2}\beta_{\bar{m}}\,,\label{equationforbeta}
\end{equation}
which is valid in the interior of the region connected to the Gaussian fixed point, below the singularity line of the equation:
\begin{equation}
1+\bar{m}^{2\eta}-2\lambda \iota(d)=0\,,
\end{equation}
and where we introduced the $\beta$-function for dimensionless mass $\bar{m}^{2\eta}:={m}^{2\eta}\Lambda^{-2\eta}$, i.e. $\Lambda^{2\eta}\beta_{\bar{m}}:=\beta_m-2\eta m^{2\eta}$.
\end{proposition}
\textit{Proof.} From RGE for $\Gamma^{(2)}(\vec{0}\,)$, $\partial\Gamma^{(2)}/\partial p_1^{2\eta} (\vec{0}\,)$ and $\Gamma^{(4)}_{\vec{0},\vec{0},\vec{0},\vec{0}}\equiv\gamma^{(4)}$, and taking into account the relations coming from Ward identities, we deduce the following relations:
\begin{align}
\bigg(\frac{\partial}{\partial t}+\beta \frac{\partial}{\partial \lambda}+\beta_m \frac{\partial}{\partial m^{2\eta}}-\gamma\bigg)&(m^{2\eta}-d\times \tau(0))=0\\
\bigg(\frac{\partial}{\partial t}+\beta \frac{\partial}{\partial \lambda}+\beta_m \frac{\partial}{\partial m^{2\eta}}-\gamma\bigg)&(1-\tau^\prime(0))=0\\
\bigg(\frac{\partial}{\partial t}+\beta \frac{\partial}{\partial \lambda}+\beta_m \frac{\partial}{\partial m^{2\eta}}-2\gamma\bigg)&\lambda Z_\lambda=0\,.
\end{align}
The two first equations are explicitly written as (we use the notation $\tau^\prime$ for $\tau^\prime(0)$ and $\tau$ for $\tau(0)$):
\begin{equation}
d \frac{\partial \tau}{\partial t}+d\,\beta \frac{\partial \tau}{\partial \lambda}-\beta_m \left(1-d\,\frac{\partial \tau}{\partial m^{2\eta}}\right)+\gamma(m^{2\eta}-d\times \tau)=0\,,
\end{equation}
\begin{equation}
\frac{\partial \tau^\prime}{\partial t}+\beta \frac{\partial \tau^\prime}{\partial \lambda}+\beta_m\frac{\partial \tau^\prime}{\partial m^{2\eta}}+\gamma(1-\tau^\prime)=0\,.
\end{equation}
For the third equations, we use the fact that $Z_\lambda=(1-\tau^\prime)/(1-2\lambda\mathcal{L}\partial\tau)$, from which it follows:
\begin{equation}
- \lambda\gamma+\beta +\frac{2\lambda \iota(d)}{1-2\lambda\mathcal{L}\partial\tau}\left(\frac{\partial}{\partial t}+\beta \frac{\partial}{\partial \lambda}+\beta_m \frac{\partial}{\partial m^{2\eta}}\right) \frac{\lambda}{1+\bar{m}^{2\eta}} =0\,.
\end{equation}
Then after a few simplifications,
\begin{equation}
-\gamma \lambda +\beta +\frac{2\lambda \iota(d)}{1+\bar{m}^{2\eta}-2\lambda \iota(d)}\left(\beta-\frac{\lambda}{1+\bar{m}^{2\eta}}\beta_{\bar{m}}\right)=0\,,\label{betastep}
\end{equation}
we then deduce the proposition assuming that $1+\bar{m}^{2\eta}-2\lambda \iota(d)\neq 0$ and $\bar{m}^{2\eta}\neq -1$.
\begin{flushright}
$\square$
\end{flushright}
As direct consequences of this statement, and from inspection of the equation \eqref{betastep}, we have:

\begin{corollary}\label{cor1}
In the deep UV limit, and the melonic sector, any fixed point $\beta=\beta_m=0$ has to satisfy (at least) one of the two conditions:
\begin{enumerate}
\item $\lambda=0$\,,
\item $\gamma=0$\,.
\end{enumerate}
\end{corollary}
The first one corresponds to the Gaussian fixed point and has no real interest at this stage. We then expect that only the second one will be of relevant interest for non-Gaussian fixed point investigations. Moreover, from equation \eqref{equationforbeta}, substituting $\beta$ and solving the resulting equations for $\beta_m$ and $\gamma$, we get:
\begin{corollary}
The $\beta$-function for mass and the anomalous dimensions be expressed only in terms of $\tau$, $\tau^\prime$, $\lambda$, $m^{2\eta}$ and $\Lambda$ as:

\begin{align}
\beta_{\bar{m}}=\,-\frac{2\eta (m^{2\eta}-d\tau)-\gamma \left(m^{2\eta}-d\left( \tau-\lambda\left(1-\frac{2\lambda\iota(d)}{1+\bar{m}^{2\eta}}\right)\frac{\partial \tau}{\partial \lambda}\right)\right) }{\Lambda^{2\eta}-d\,\left(\Lambda^{2\eta}\frac{\partial \tau}{\partial m^{2\eta}}+\frac{2\lambda^2\iota(d)}{1+\bar{m}^{2\eta}}\frac{\partial \tau}{\partial \lambda}\right)}\,,\label{eqbetam2}
\end{align}

\begin{eqnarray}
\gamma=\frac{2\eta (m^{2\eta}-d\tau) \,\Omega}{1- \tau^\prime+\lambda\left(1-\frac{2\lambda\iota(d)}{1+\bar{m}^{2\eta}}\right)\frac{\partial \tau^\prime}{\partial \lambda}+\left({m^{2\eta}-d\left( \tau-\lambda\left(1-\frac{2\lambda\iota(d)}{1+\bar{m}^{2\eta}}\right)\frac{\partial \tau}{\partial \lambda}\right)}\right)\,\Omega}\,,\label{eqgamma2}
\end{eqnarray}
with:
\begin{equation}
\Omega:=\frac{\Lambda^{2\eta}\frac{\partial \tau^\prime}{\partial m^{2\eta}}+\frac{2\lambda^2\iota(d)}{1+\bar{m}^{2\eta}}\frac{\partial \tau^\prime}{\partial \lambda}}{\Lambda^{2\eta}-d\,\left(\Lambda^{2\eta}\frac{\partial \tau}{\partial m^{2\eta}}+\frac{2\lambda^2\iota(d)}{1+\bar{m}^{2\eta}}\frac{\partial \tau}{\partial \lambda}\right)}
\end{equation}
\end{corollary}
The set of three equations, \eqref{equationforbeta}, \eqref{eqbetam2} and \eqref{eqgamma2} show explicitly that the knowledge of $\tau$ and $\tau^\prime$ determine entirely the behavior of the RG flow in the deep UV.

\section{Large $d$ behavior of the Feynman amplitudes}\label{section3}

In this section, we investigate the large rank behavior of the melonic Feynman amplitudes. We start with the heuristic computation of the relevant quantities $\tau$ and $\tau^\prime$, and then extend our results to all orders of the perturbative expansion, for two-point and vacuum amplitudes. A second time, we build an exact renormalization group equation and show that no fixed point may be found in this large $d$-limit.
The result of this section, therefore, solves the closed equation of the two-point correlation function at large rank limit exploration.

\subsection{One and two-loops investigation}
i)\,\, \textit{One-loop computation.} A typical leading order contribution to the one-loop 1PI two-point function has been drawn on Figure \ref{fig2} -- on left. Note that there are two Wick contractions for this configuration, meaning that for each melonic vertex, the contribution of the diagram has to be counted twice. From the Feynman rules, we then deduce the one-loop self-energy as:
\begin{align}\label{susat2}
\tau^{(1)}(p_i)=\sum_{\vec{q}\in\mathbb{Z}^{d-1}} \frac{\Theta(\Lambda^{2\eta}-\vec{q}\,^{2\eta}-p_i^{2\eta})}{\vec{q}\,^{2\eta}+p_i^{2\eta}+m^{2\eta}}\,.
\end{align}
where the subscript $(1)$ refers to the number of loops. Due to the large $\Lambda$ limit, we can simplify the computation by taking the continuum limit and replacing the sum with an integral, without consequences on the leading order contributions. We introduce the continuous variables $x_i:= p_i/\Lambda$. Then the equation \eqref{susat2} becomes:
\begin{equation}
\tau^{(1)}(\Lambda\, x)=-2\lambda \,\Lambda^{d-1-2\eta}\, \int d\textbf{x} \delta_{\Lambda}(x_1-x) \frac{\Theta(1-\vec{\textbf{x}}\,^{2\eta})}{\vec{\textbf{x}}\,_\bot^{2\eta}+x^{2\eta}+\bar{m}^{2\eta}}\,,
\end{equation}
where we introduced the dimensionless mass $\bar{m}^{2\eta}=m^{2\eta}/\Lambda^{2\eta}$, the $\delta$-distribution of size $1/\Lambda$ : $\delta_{\Lambda}(x_1-x):=\Lambda\delta_{pq_1}$, $d\textbf{x}:=\prod_i dx_i$, and $\textbf{x}_\bot\equiv \textbf{x}_{\bot\,1}=(x_2,\cdots,x_d)$. Defining the continuous function ${\tau}$ as ${\tau}(x):= \tau(\Lambda\, x)/\Lambda^{2\eta}$, and because $d-1-2\eta=2\eta$, we get finally:
\begin{equation}
{\tau}^{(1)}(x)=-2\lambda \int d\textbf{x}_\bot \frac{\Theta(1-\vec{\textbf{x}}\,^{2\eta})}{\vec{\textbf{x}}\,_\bot^{2\eta}+x^{2\eta}+\bar{m}^{2\eta}}\,,\label{equationnum}
\end{equation}
where we have formally took the $\Lambda\to \infty$ limit. This integral may be computed from the results given in Appendix \ref{App2}, and we deduce the explicit formula:
\begin{equation*}
\tau^{(1)}(p_i)=-\,2^{d}\lambda( \Lambda^{2\eta}-p^{2\eta}_i) \bigg[\Gamma\left(\frac{d+1}{d-1}\right)\bigg]^{d-1}
\times\bigg[1-\frac{m^{2\eta}+p^{2\eta}_i}{ \Lambda^{2\eta}-p^{2\eta}_i}\ln\left(\frac{\Lambda^{2\eta}+m^{2\eta}}{m^{2\eta}+p^{2\eta}_i}\right)\bigg]\,,
\end{equation*}
From this we get:
\begin{equation}
\Sigma^{(1)}(\vec{p}=\vec{0})=-2^{d}d\lambda\Lambda^{2\eta} \bigg[\Gamma\left(\frac{d+1}{d-1}\right)\bigg]^{d-1}\bigg[1-\bar{m}^{2\eta}\ln\left(\frac{1+\bar{m}^{2\eta}}{\bar{m}^{2\eta}}\right)\bigg]\,,\label{masscorr}
\end{equation}
and:
\begin{equation}
\tau^{(1)\,\prime}(0)=2^{d}\lambda \bigg[\Gamma\left(\frac{d+1}{d-1}\right)\bigg]^{d-1}(1+\bar{m}^{2\eta})\ln\left(\frac{1+\bar{m}^{2\eta}}{\bar{m}^{2\eta}}\right)\,.
\end{equation}
\\
ii)\,\,\textit{Two-loops computation.} We now move on to the two loops computation of the self-energy $\tau^{(2)}(p)$. At the leading order in the deep UV, there is only one relevant diagram, which is:
\begin{equation}
\tau(p)=\sum_{i=1}^d\,\,\vcenter{\hbox{\includegraphics[scale=0.8]{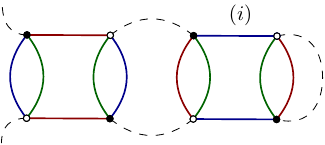} }}\,.
\end{equation}
We recall that we need to compute only $\tau(0)$ and $\tau^\prime(0)$ to build the renormalization group flow in the deep UV. From the previous diagram, it is quite natural to split the computation as the sum of two distinct contributions:
\begin{align}
\tau^{(2)}(p)=\tau^{(2)\,\parallel}(p)+(d-1)\tau^{(2)\,\perp}(p)\,.
\end{align}
The first term, which we denoted as $\tau^{(2)\,\parallel}(p)$ corresponds to the configuration where the two vertices are the same (same color); and the second one corresponds to the case where the two vertices are different (different colors). We will compute each term separately. The Feynman rules, it follows:
\begin{equation}
\tau^{(2)\,\parallel}(p)=s_\parallel \lambda^2\times \sum_{\vec{q}_\bot\,,\vec{k}_\bot\in \mathcal{D}(p)} \frac{1}{(\vec{q}\,^{2\eta}_{\bot}+p^{2\eta}+m^{2\eta})^2}\frac{1}{\vec{k}\,^{2\eta}_{\bot}+p^{2\eta}+m^{2\eta}}\,,
\end{equation}
where $s_\parallel$ is a symmetry factor and $$\mathcal{D}(p):=\{\vec{q}_\bot\in \mathbb{Z}^{d-1}\vert \vec{q}\,^{2\eta}_\bot\leq \Lambda^{2\eta}-p^{2\eta}\}.$$
The symmetry factor receives two contributions. First, we have a factor $1/2!$ coming from the expansion of the exponential. Secondly, the number of allowed contractions leading to such a melonic diagram can be given by the following. There is a first factor $2$ coming from the choice of the vertex on which the external edges are hooked, and a second factor $2$ coming from the orientation of the vertex. Finally, a third factor $2$ arises from the orientation of the second vertex (there are two different ways to hook this vertex to the first one, and a single possibility to create the last internal line of length one). As a result:
\begin{equation}
s_\parallel=\frac{1}{2!}\times 2\times 2\times 2=4\,.
\end{equation}
At this stage, we use the sums $S_1$ and $S_2$ defined in the Appendix \ref{App2}. Indeed, $\tau^{(2)\,\parallel}(p)$ can be factorized in two contributions corresponding to the two sub-melonic diagrams with two and four external points:
\begin{equation*}
\tau^{(2)\,\parallel}(p)=4 \lambda^2\, \left[ \sum_{\vec{q}_\bot \in \mathcal{D}(p)} \frac{1}{(\vec{q}\,^{2\eta}_{\bot}+p^{2\eta}+m^{2\eta})^2}\right]\times \left[ \sum_{\vec{k}_\bot \in \mathcal{D}(p)} \frac{1}{\vec{k}\,^{2\eta}_{\bot}+p^{2\eta}+m^{2\eta}}\right]\,,
\end{equation*}
Then using the sums, \eqref{formulaK} and \eqref{formulaL} in the Appendix \ref{App2} we get
\begin{align}\label{taupara}
\nonumber\tau^{(2)\,\parallel}(p)=4[\iota(d)]^2 (\Lambda^{2\eta}-p^{2\eta})&\lambda^2\, \bigg[1-\frac{m^{2\eta}+p^{2\eta}}{ \Lambda^{2\eta}-p^{2\eta}}\ln\left(\frac{\Lambda^{2\eta}+m^{2\eta}}{m^{2\eta}+p^{2\eta}}\right)\bigg]\\
&\times \left[\ln\left(\frac{\Lambda^{2\eta}+m^{2\eta}}{m^{2\eta}+p^{2\eta}}\right)-\frac{\Lambda^{2\eta}-p^{2\eta}}{\Lambda^{2\eta}+m^{2\eta}} \right]\,.
\end{align}
The computation of the quantity $\tau^{(2)\,\bot}$ may be given easily, due to the overlapped momentum between the two loops of the diagram. We get:
\begin{equation}
\tau^{(2)\,\bot}(p)=s_\bot \lambda^2\times \sum_{\vec{q}_\bot\in \mathcal{D}(p)\,,\vec{k}_\bot\in \mathcal{D}(q_2)} \frac{1}{(\vec{q}\,^{2\eta}_{\bot}+p^{2\eta}+m^{2\eta})^2}\frac{1}{\vec{k}\,^{2\eta}_{\bot}+q^{2\eta}_2+m^{2\eta}}\,.\label{twoloop2}
\end{equation}
By considering the equation \eqref{Feynman} and for simplicity, we will compute separately $\tau^{(2)\,\bot}(0)$ and $\tau^{(2)\,\bot\,\prime}(p)$. We get, in the continuum limit:
\begin{align}
\nonumber\tau^{(2)\,\bot}(0)=2s_\bot \lambda^2\Lambda^{2\eta}& \int d\vec{q}_\bot d\vec{k}_\bot \Theta(1-\vec{q}\,_\bot^{2\eta})\Theta(1-\vec{k}_{\bot}^{2\eta}-q_2^{2\eta})\\
&\qquad \times \int_0^1 du_1du_2 \frac{u_1\delta(1-u_1-u_2) }{(u_1\vec{q}\,^{2\eta}_{\bot^\prime}+u_2\vec{k}\,^{2\eta}_{\bot}+q_2^{2\eta}+\bar{m}^{2\eta})^3}\,,
\end{align}
where we kept the notation $q$ and $k$ for continuous variables. Now we introduce the integral representation of the Heaviside $\Theta$-functions, leading to:
\begin{align}
\nonumber\tau^{(2)\,\bot}(0)=2s_\bot \lambda^2\Lambda^{2\eta}& \int d\vec{q}_\bot d\vec{k}_\bot\int_0^1 dy_1dy_2 \delta(y_1-\vec{q}\,_\bot^{2\eta})\delta(y_2-\vec{k}_{\bot}^{2\eta}-q_2^{2\eta})\\
&\qquad \times \int_0^1 du_1du_2 \frac{u_1\delta(1-u_1-u_2) }{(u_1\vec{q}\,^{2\eta}_{\bot^\prime}+u_2\vec{k}\,^{2\eta}_{\bot}+q_2^{2\eta}+\bar{m}^{2\eta})^3}\,.
\end{align}
Due to the properties of the $\delta$-distribution, this relation takes the simple form:
\begin{align}
\nonumber\tau^{(2)\,\bot}(0)=2s_\bot \lambda^2\Lambda^{2\eta}& \int d\vec{q}_\bot d\vec{k}_\bot\int_0^1 dy_1dy_2 \delta(y_1-\vec{q}\,_\bot^{2\eta})\delta(y_2-\vec{k}_{\bot}^{2\eta}-q_2^{2\eta})\\
&\qquad \times \int_0^1 du_1du_2 \frac{u_1\delta(1-u_1-u_2) }{(u_1y_1+u_2y_2+\bar{m}^{2\eta})^3}\,.
\end{align}
We make the change of variables: $\vec{q}\,_\bot^{2\eta}\to y_1\vec{q}\,_\bot^{2\eta}$, and $\vec{k}\,_\bot^{2\eta}\to (y_2-y_1q_2^{2\eta})\vec{k}\,_\bot^{2\eta}$; splitting $\tau^{(2)\,\bot}(0)$ into two contributions:
\begin{equation}
\tau^{(2)\,\bot}(0)=2s_\bot \lambda^2\Lambda^{2\eta} [L_1(d)-L_2(d)]\,,
\end{equation}
where $L_1(d)$ and $L_2(d)$ are defined as:
\begin{align}
\nonumber L_1(d):=\int d\vec{q}_\bot d\vec{k}_\bot \delta(1-\vec{q}\,_\bot^{2\eta})&\delta(1-\vec{k}_{\bot}^{2\eta}) \times \int_0^1 du_1du_2u_1\delta(1-u_1-u_2)\\
&\quad \times \int_0^1 dy_1dy_2\frac{y_1y_2 }{(u_1y_1+u_2y_2+\bar{m}^{2\eta})^3}\,,
\end{align}
\begin{align}
\nonumber L_2(d):=\int d\vec{q}_\bot d\vec{k}_\bot \delta(1-\vec{q}\,_\bot^{2\eta})&\delta(1-\vec{k}_{\bot}^{2\eta}) \times \int_0^1 du_1du_2u_1\delta(1-u_1-u_2)\\
&\quad \times \int_0^1 dy_1dy_2\frac{y_1^2\,q_2^{2\eta} }{(u_1y_1+u_2y_2+\bar{m}^{2\eta})^3}\,.
\end{align}
Note that the role-playing by the variable $q_2$ is arbitrary. Then, we can sum over all choices of them, and finally dividing the result by $d-1$ we get:
\begin{align*}
\nonumber L_2(d):=\frac{1}{d-1}\int d\vec{q}_\bot d\vec{k}_\bot \delta(1-\vec{q}\,_\bot^{2\eta})&\delta(1-\vec{k}_{\bot}^{2\eta}) \times \int_0^1 du_1du_2u_1\delta(1-u_1-u_2)\\
&\quad \times \int_0^1 dy_1dy_2\frac{y_1^2 }{(u_1y_1+u_2y_2+\bar{m}^{2\eta})^3}\,.
\end{align*}
Interestingly, the $d$-dependence of the two loop integrals $L_1$ and $L_2$ can be factorized, (the same phenomena can be observed at one loop, as shown in the Appendix \ref{App2}). Moreover, this factorization is a consequence of the role played by our deformation parameter $\eta$. Then we choose $\eta$ such that the theory remains just-renormalizable in any dimensions, i.e. the loop structure remains the same in any dimensions. Finally $\tau^{(2)\,\bot}(0)$ takes the form:
\begin{equation}
\tau^{(2)\,\bot}(0)=2s_\bot \lambda^2 [\iota(d)]^2\Lambda^{2\eta} \left[R_1-\frac{1}{d-1}R_2\right]\,,\label{eqtwotau}
\end{equation}
where:
\begin{equation}
R_1= \int_0^1 du_1du_2u_1\delta(1-u_1-u_2) \times \int_0^1 dy_1dy_2\frac{y_1y_2 }{(u_1y_1+u_2y_2+\bar{m}^{2\eta})^3}\,,
\end{equation}
and:
\begin{equation}
R_2= \int_0^1 du_1du_2u_1\delta(1-u_1-u_2) \times \int_0^1 dy_1dy_2\frac{y_1^2 }{(u_1y_1+u_2y_2+\bar{m}^{2\eta})^3}\,.
\end{equation}
The first integral may be straightforwardly computed: $R_1$ is nothing but the same contribution in the final expression of $\tau^{(2)\,\parallel}(p=0)$ given in \eqref{taupara}:
\begin{equation}
R_1= \frac{1}{2}\bigg[1-\frac{m^{2\eta}}{ \Lambda^{2\eta}}\ln\left(\frac{\Lambda^{2\eta}+m^{2\eta}}{m^{2\eta}}\right)\bigg]\times \left[\ln\left(\frac{\Lambda^{2\eta}+m^{2\eta}}{m^{2\eta}}\right)-\frac{\Lambda^{2\eta}}{\Lambda^{2\eta}+m^{2\eta}} \right]\,,
\end{equation}
Moreover, it is easy to check that $s_\bot=4$. Indeed, with respect to the previous counting for $s_\parallel$, we lack a factor $2$ coming from the exchange of the vertices, but we have an additional factor $2$ arising from the expansion of the square of the interaction, which concerns only the contributions with vertices of different colors. Finally, the $2$-loops contribution to $\tau(p=0)$ writes as:
\begin{equation}
\tau^{(2)}(p=0)=8\,d\,[\iota(d)]^2 \Lambda^{2\eta}\lambda^2 \left[\, R_1+\frac{1}{d-1}R_2\,\right]\,. \label{twolooptau}
\end{equation}
The last term $R_2$ can be interpreted as an overlapping effect and in the large $d$ limit this quantity disappears.
\\

\noindent
The first derivative $\partial\tau^{(2)\,\bot}/\partial p^{2\eta}$ for zero momentum can be derived follows the same strategy. From expression \eqref{twoloop2}, we get :
\begin{align}
\nonumber\frac{\partial \tau^{(2)\,\bot}}{\partial p^{2\eta}}(p=0)=&-8 \lambda^2\times \int d\vec{q}_\bot d\vec{k}_\bot \frac{\Theta(1-\vec{q}\,_\bot^{2\eta})}{(\vec{q}\,^{2\eta}_{\bot}+\bar{m}^{2\eta})^3}\frac{\Theta(1-\vec{k}_{\bot}^{2\eta}-q_2^{2\eta})}{\vec{k}\,^{2\eta}_{\bot}+q^{2\eta}_2+\bar{m}^{2\eta}}\\
&- 4\lambda^2\times \int d\vec{q}_\bot d\vec{k}_\bot \frac{\delta(1-\vec{q}\,_\bot^{2\eta})}{(\vec{q}\,^{2\eta}_{\bot}+\bar{m}^{2\eta})^2}\frac{\Theta(1-\vec{k}_{\bot}^{2\eta}-q_2^{2\eta})}{\vec{k}\,^{2\eta}_{\bot}+q^{2\eta}_2+\bar{m}^{2\eta}}\cr
=&P_1+P_2\,.\label{twoloopZ}
\end{align}
where
\begin{align}
\nonumber P_2&=- 4\lambda^2\times\int d\vec{q}_\bot d\vec{k}_\bot \frac{\delta(1-\vec{q}\,_\bot^{2\eta})}{(1+\bar{m}^{2\eta})^2}\frac{\Theta(1-\vec{k}_{\bot}^{2\eta}-q_2^{2\eta})}{\vec{k}\,^{2\eta}_{\bot}+q^{2\eta}_2+\bar{m}^{2\eta}}\\\nonumber
&= - 4\lambda^2\times \int_0^1 dy\int d\vec{q}_\bot d\vec{k}_\bot \frac{\delta(1-\vec{q}\,_\bot^{2\eta})}{(1+\bar{m}^{2\eta})^2}\frac{\delta(y-\vec{k}_{\bot}^{2\eta}-q_2^{2\eta})}{\vec{k}\,^{2\eta}_{\bot}+q^{2\eta}_2+\bar{m}^{2\eta}}\\
&= - \frac{4\lambda^2}{(1+\bar{m}^{2\eta})^2}\times\int d\vec{q}_\bot d\vec{k}_\bot \int_0^1 dy\frac{y-q_2^{2\eta}}{y+\bar{m}^{2\eta}}\delta(1-\vec{q}\,_\bot^{2\eta}) \delta(1-\vec{k}_{\bot}^{2\eta})\,.
\end{align}
Then by summing all the possibles choices of the variable $q2$ and dividing by $d-1$, we get:
\begin{equation}
P_2= - \frac{4\lambda^2}{(1+\bar{m}^{2\eta})^2}\,[\iota(d)]^2 \int_0^1 dy\frac{y-\frac{1}{d-1}}{y+\bar{m}^{2\eta}}\,.
\end{equation}
This expression corresponds to the computation of the effective mass correction and in large $d$, and we retain:
\begin{equation}
P_2\underset{d\gg1}{\longrightarrow}- \frac{4\lambda^2}{(1+\bar{m}^{2\eta})^2}\,[\iota(d)]^2 \left(1-\bar{m}^{2\eta}\ln\left(\frac{1+\bar{m}^{2\eta}}{\bar{m}^{2\eta}}\right)\right)\,.\label{P2}
\end{equation}
In the same manner:
\begin{equation}
P_1\underset{d\gg1}{\longrightarrow}-8 \lambda^2[\iota(d)]^2\times \int_0^1\frac{y_1dy_1}{(y_1+\bar{m}^{2\eta})^3}\frac{y_2dy_2}{y_2+\bar{m}^{2\eta}}\,,
\end{equation}
and after integration we get
\begin{align}
\nonumber P_1\underset{d\gg1}{\longrightarrow}-8 \lambda^2[\iota(d)]^2\times& \left(\frac{1}{1+\bar{m}^{2\eta}}\left(\frac{1}{2}\frac{\bar{m}^{2\eta}}{1+\bar{m}^{2\eta}}-1\right)+\frac{1}{2}\frac{1}{\bar{m}^{2\eta}}\right)\\
&\qquad\qquad\times\left(1-\bar{m}^{2\eta}\ln\left(\frac{1+\bar{m}^{2\eta}}{\bar{m}^{2\eta}}\right)\right)\,.\label{P1}
\end{align}
Note that, as the one-loop correction, the two loops function is not perturbative in $\bar{m}^{2\eta}$. To summarize, in the large $d$ limit, we get for two loops contributions to $\tau$ and $\tau^\prime$ :
\begin{equation}
\tau^{(2)}=-4d [\iota(d)]^2 \lambda^2\Lambda^{2\eta} (1+\ln(\bar{m}^{2\eta}))\,,
\end{equation}
\begin{equation}
\tau^{(2)\,\prime} =4d [\iota(d)]^2 \lambda^2 \left(1-\frac{1}{\bar{m}^{2\eta}}\right)\,.
\end{equation}
As remark, the infrared divergences that we observe at two loops order occur in the computation at $n$-loops and these divergences are increased with the number $n$ of loops as
\begin{equation}
\int_0^\Lambda \frac{d\vec{p}}{(\vec{p}\,^2)^n}\equiv \Lambda^{d-2n}\,.
\end{equation}

\subsection{Structure of the n-loops graphs in large $d$ limit}\label{nloop}
In this section, we extend the result of the previous section to arbitrary large Feynman graphs in the large $d$ limit; providing the first hard statement of this paper. Heuristically, if we discard the terms mixing coupling and mass, for $n$--loops, the expected following behavior for $\tau(p)$:
\begin{equation}
\tau^{(n)}(p)\propto\, d^{n-1} \,(2\iota(d))^n \times \lambda^{n}\,.\label{stat3}
\end{equation}
This behavior can be proved recursively but has proved for $n=1$ and $n=2$ which highlight the initial origin of the different factors in \eqref{stat3}. For instance, a factor $\iota(d)$ seems to be associated with each loop. The origin of the factor $d$ moreover is clear. As recalled in the Appendix \ref{App1}, the leading order contributions are trees in the so-called \textit{intermediate field representation}, then, the typical graphs contributing to $\tau^{(p)}$ and $\tau^{(p)\,\prime}$ are trees with $p$ colored edges. All the edges are color-free, except the color of the edge corresponding to the single boundary vertex. As a result, there are $d^{p-1}$ different trees with the same uncolored combinatorial structure; and the cardinality of $\tau^{(p)}$ and $\tau^{(p)\,\prime}$ is $d^{p-1}$ times a purely combinatorial number depending only on $p$. In this subsection, we will prove this intuition, and investigate the structures and properties of higher-loops diagrams in the large $d$ limit. More precisely, we will prove the following statement:
\begin{proposition}\label{propositionbehavior}
Let $\mathcal{T}_n$ be a $2$-points $n$-loops tree contributing to $\tau^{(n)}(p)$ and let $r$ be its root loop vertex, at which the external colored edge is hooked. In the UV sector ($\Lambda\gg1$) and in the large dimension limit ($d\gg n\geq 1$), the perturbative $n$-loops amplitude $\mathcal{A}_{\mathcal{T}_n}$ behaves in $\lambda$ and $d$ like:
\begin{equation}
\mathcal{A}_{\mathcal{T}_n}(p)=(\Lambda^{2\eta}-p^{2\eta}) \,c_n(\bar{m}^{2\eta},p)\,(\iota(d))^n\,,
\end{equation}
where $c_n(\bar{m}^{2\eta},p)$ includes a proper mass and external momenta dependence:
\begin{equation}
c_n(\bar{m}^{2\eta})=(-1)^{n-1}\left[\prod_{b\in\mathcal{T}_n/r}\frac{\omega^{(m(b))}(\bar{m}^{2\eta})}{[(m(b)-1)!]} \right]\times \mathcal{A}_r(p),
\end{equation}
where $\mathcal{T}_n\in\mathbb{T}_n$ and $\mathbb{T}_n$ denotes the set of trees with $n$ loop vertices and different colors on their edges, $m(b)$ is the coordination number at the loop-vertex $b$, and $\omega^{(m(b))}(\bar{m}^{2\eta})$ is the $m(b)$--th derivative of $\omega$ defined as:
\begin{align}
\nonumber\omega(\bar{m}^{2\eta})&:=\frac{1}{2}\left(\ln(1+\bar{m}^{2\eta})+\bar{m}^{2\eta}-(\bar{m}^{2\eta})^2 \ln\left(\frac{1+\bar{m}^{2\eta}}{\bar{m}^{2\eta}}\right)\right)\\
&\,\,\equiv \int_0^{\bar{m}^{2\eta}}dx\int_0^1dy\frac{y}{y+x}\,.
\end{align}
Finally, the root amplitude $\mathcal{A}_r(p)$ sharing the external momenta dependence writes as:
\begin{equation}
\mathcal{A}_r(p):=\frac{1}{(m(r)-1)!}\frac{\partial^{m(r)-1}}{\partial (\bar{m}^{2\eta})^{m(r)-1}}
\,\bigg[1-\frac{\bar{m}^{2\eta}+\frac{p^{2\eta}}{\Lambda^{2\eta}}}{1-\frac{p^{2\eta}}{\Lambda^{2\eta}}}\ln\left(\frac{1+\bar{m}^{2\eta}}{\bar{m}^{2\eta}+\frac{p^{2\eta}}{\Lambda^{2\eta}}}\right)\bigg].
\end{equation}
\end{proposition}
\textit{Proof.} The statement has been proved for $n=1$ and $n=2$. Let us then provide the general proof by recurrence, i.e. we assume that the proposition holds for $n$ loops, and we will prove that the expected structure survives for $n+1$ loops. From proposition \ref{theoremtreees} (see Appendix \ref{App1}), the single colored self-energy $\tau^{(n)}(p)$ of order $n$, may be written as a sum of rooted trees with $n$ loop-vertices in the intermediate field representation. The root is a colored edge hooked to one of the loop vertices, which we call the external loop vertex, for instance:
\begin{equation}
\mathcal{T}_8\,\equiv\,\vcenter{\hbox{\includegraphics[scale=1]{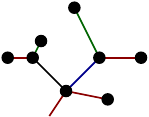} }}\,,
\end{equation}
is such a typical tree, with a root of color red. \\

\noindent
Now, let $\mathbb{T}_n$ be the set of such a trees with $n$ loop vertices, and $F$ be a surjective map from $\mathbb{T}_{n}$ to $\mathbb{T}_{n+1}$:
\begin{equation}
F:\mathbb{T}_{n}\to \mathbb{T}_{n+1}\,.
\end{equation}
The map $F$ can be constructed explicitly. Indeed, for any tree $\mathcal{T}_{n+1}\in \mathbb{T}_{n+1}$ it is not hard to check that there exists a single tree in $\mathbb{T}_{n}$ such that $\mathcal{T}_{n+1}$ may be obtained from $\mathcal{T}_{n}$ by adding one leaf:
\begin{equation}
\vcenter{\hbox{\includegraphics[scale=1]{recuropen1.pdf} }}\quad \to \quad\vcenter{\hbox{\includegraphics[scale=1]{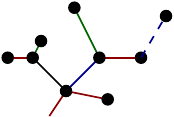} }} \,.
\end{equation}
We can then consider $F$ as the transformation sending any tree $\mathcal{T}_n\in\mathbb{T}_n$ to a set $F[\mathcal{T}_n]\subset \mathbb{T}_{n+1}$ of cardinality $(d-n)\times n$ whose elements are any trees with $n+1$ loop vertices obtained from $\mathcal{T}_n$ by adding a leaf.
Moreover, we expect that $F[\mathcal{T}_n]\cap F[\mathcal{T}_n^\prime]\neq \emptyset$ in general, because any tree in $\mathbb{T}_{n+1}$ have more than one antecedent in $\mathbb{T}_{n}$. As an illustration, the tree:
\begin{equation}
\includegraphics[scale=1]{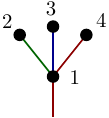}
\end{equation}
with $n=4$ has three antecedents, obtained from the deletion of one among the three leaves hooked to the loop-vertex labeled $1$:
\begin{equation}
\includegraphics[scale=0.8]{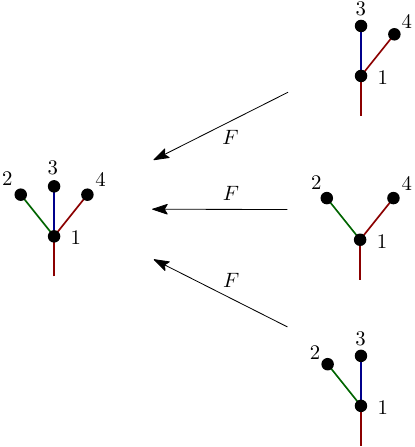}
\end{equation}
ensuring the surjectivity of the map $F$. \\
\noindent
We have now all the materials to build our recurrence. Let us consider a tree $\mathcal{T}_n$, with $n$ loop-vertex, to which we add a leaf $\mathcal{L}_{c}$ of color $c$. We denote by $\mathcal{T}_n\ast_b \mathcal{L}_{c}$ the resulting tree with $n+1$ loop-vertices; $b$ being the loop-vertex at which the leaf is hooked. We assume that $c\neq 1$, where $1$ refers to the color of the root. As we have seen for the computation of the two-loops $2$-point function, this restriction does not affect the large $d$ limit, the color $c$ being chosen among $d-1$ colors rather than $d$. From Feynman rules, the amplitude $\mathcal{A}_{\mathcal{T}_n\ast \mathcal{L}_{c}}(p)$ for the resulting $n+1$ graph can be written explicitly as:
\begin{equation}
\mathcal{A}_{\mathcal{T}_n\ast_b \mathcal{L}_{c}}(p)=\sum_{q\in\mathbb{Z}} \mathcal{A}_{\mathcal{T}_n^\prime}(p,q) \mathcal{A}_{ \mathcal{L}_{c}}(q)\,,
\end{equation}
where $\mathcal{A}_{ \mathcal{L}_{c}}(q)$ is the Feynman amplitude for the leaf $ \mathcal{L}_{c}$ and $\mathcal{T}_n^\prime$ the $2$-root tree obtained from the single root tree $\mathcal{T}_n$ by hooking an half edge of color $c$ to the vertex $b$:
\begin{equation}
\mathcal{T}_{n+1}=\vcenter{\hbox{\includegraphics[scale=1]{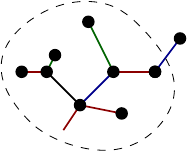} }}\quad \to\quad \mathcal{T}_n^\prime=\vcenter{\hbox{\includegraphics[scale=1]{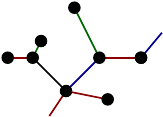} }}\,.
\end{equation}
The relation between $\mathcal{A}_{\mathcal{T}_n}$ and $\mathcal{A}_{\mathcal{T}_n^\prime}$ can be obtained as follow. From Feynman rules, we get the explicit expression for $\mathcal{A}_{\mathcal{T}_n}$ as:
\begin{equation}
\mathcal{A}_{\mathcal{T}_n}(p)=\prod_{b^\prime\in \mathbb{B}} \sum_{\{\vec{p}_{b^\prime}\}}C^{l(b^\prime)}(\vec{p}_{b^\prime})\prod_{e\in \partial f_{e\,r}}\delta_{p_{s(e)}p_{t(e)}}\delta_{p_{s(e)}p}\prod_{f\in \mathbb{F}} \prod_{e\in \partial f}\delta_{p_{s(e)}p_{t(e)}}\,,
\end{equation}
where $\mathbb{B}$ and $\mathbb{F}$ are respectively the sets of loop vertices and internal faces, and $\partial f$ is the subset of mono-colored edges building the colored face $f$. Moreover, $s(e)$ and $t(e)$ denote respectively the source and target loop vertices bounding the edge $e$, and $f_{e\,r}$ is the external face running through the root. A \textit{colored face} on a tree corresponds to a colored and unclosed path, for internal as for external ordinary faces, passing through loop vertices at which are hooked some connected components(see Figure \ref{figface} below).

\begin{figure}
\begin{center}
\includegraphics[scale=0.7]{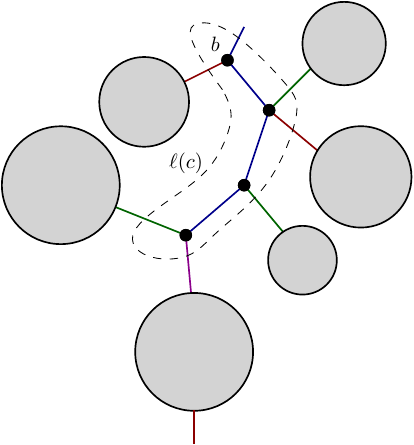}
\captionof{figure}{The colored path corresponds to an external blue face, and the connected components are hooked to the loop vertices along the path. }\label{figface}
\end{center}
\end{figure}

\noindent
Let $f_c$ be the external face of color $c$ that we created on $\mathcal{T}_n^\prime$, starting at the loop-vertex $b$, and $\ell(c)$ the corresponding colored path, having $b$ as a boundary. On $\mathcal{T}_n^\prime$, in addition to our external edge of color $c$, we have $m(b)-1$ colored edges hooked to $b$. $m(b)$ is the coordination number of the vertex $b$ on $\mathcal{T}_n$; therefore $m(b)\leq n$ is a trivial bound. Each of these colored edges is hooked to connected components, like in Figure \ref{figface}, where $m(b)=2$. The number of independent configurations, that is, the number of different choices for the $m(b)$ colors is nothing but the counting of the number of different manners to choose $m(b)$ colors among $d$. More precisely, let $m_{c^\prime}(b)$ be the number of colored edges hooked to $b$ with color $c^\prime$. The number of manners to choose these $m(b)$ edges is then:
\begin{equation}
\frac{m(b)!}{\prod_{c^\prime}^{c(b)} m_{c^\prime}(b)!}\times \frac{d!}{c(b)!(d-c(b))!}\,,
\end{equation}
where $c(b)\leq m(b)$ is the number of different colors for edges hooked to $b$. For large $d$, and from the standard Stirling formula, we get:
\begin{align}
\nonumber \frac{d!}{c(b)!(d-c(b))!} &\sim \frac{1}{c(b)!}\frac{1}{(1-c(b)/d)^{d-c(b)+1}} \left(\frac{d}{e}\right)^{c(b)} \to \frac{1}{c(b)!} \left(\frac{d}{e}\right)^{c(b)}\,.
\end{align}
The distribution is stitched for $c(b)=m(b)$; and a little deviation from this configuration receives a weight $1/d$. For instance, the first deviation: $c(b)=m(b)-1$ arise with relative weight: $m(b) (d/e)^{-1} \leq n (d/e)^{-1}$. Then in the limit $d\gg n \gg 1$, this term remains small and is crushed by the dominant configuration, ensuring that $\ell(c)$ must have zero length for the dominant configurations. In other words, our added leaf on $b$ opens an internal ordinary face of length one or more that is very small. As a result, the structure of $\mathcal{T}_n$ that we have to consider is the following:
\begin{equation}
\mathcal{A}_{\mathcal{T}_n}(p)=\sum_{\vec{p}_b} C^{c(b)}(\vec{p}_b) \prod_{m=1}^{c(b)} \mathcal{A}_{\mathcal{T}_{p(m)}}(p_{c(m)})\,,
\end{equation}
where $\vec{p}_b$ is the internal momenta running through the loop vertex $b$, $p(m)$ designates the order of the connected component $\mathcal{T}_{p(m)}$ and $c(m)$ the color of the edge $m$. Note that, because $\mathcal{T}_n$ is a two-point graph, one of the connected components has to be a four points graph. Note that to simplify the notations, we only indicate the external variables for edges hooked to the vertex $b$. In the same way, the expression for $\mathcal{A}_{\mathcal{T}_{n+1}}$ becomes:
\begin{equation}
\mathcal{A}_{\mathcal{T}_{n+1}}(p)=\sum_{\vec{q}}C(\vec{q}\,)\,\left[\sum_{\vec{p}_b} \delta_{p_c\,q_c}C^{c(b)+1}(\vec{p}_b) \prod_{m=1}^{c(b)} \mathcal{A}_{\mathcal{T}_{p(m)}}(p_{c(m)})\right] \,,
\end{equation}
or, more explicitly:
\begin{equation*}
\mathcal{A}_{\mathcal{T}_{n+1}}(p)=\sum_{\vec{q}}\frac{\Theta(\Lambda^{2\eta}-\vec{q}\,^{2\eta})}{\vec{q}\,^{2\eta}+m^{2\eta}}\,\left[\sum_{\vec{p}_b} \delta_{p_c\,q_c}\frac{\Theta(\Lambda^{2\eta}-\vec{p}_b\,^{2\eta})}{(\vec{p}_b\,^{2\eta}+m^{2\eta})^{c(b)+1}} \prod_{m=1}^{c(b)} \mathcal{A}_{\mathcal{T}_{p(m)}}(p_{c(m)})\right] \,.
\end{equation*}
In the continuum limit, this expression becomes:
\begin{align}
\nonumber\mathcal{A}_{\mathcal{T}_{n+1}}(p)=\Lambda^{2\eta}\int_{\vec{q}}\frac{\Theta(1-\vec{q}\,^{2\eta})}{\vec{q}\,^{2\eta}+\bar{m}^{2\eta}}&\,\bigg[\int_{\vec{p}_b} \delta_{p_c\,q_c}\frac{\Theta(1-\vec{p}_b\,^{2\eta})}{(\vec{p}_b\,^{2\eta}+\bar{m}^{2\eta})^{c(b)+1}}\times \prod_{m=1}^{c(b)}\bar{ \mathcal{A}}_{\mathcal{T}_{p(m)}}(p_{c(m)})\bigg] \,,
\end{align}
where we recall that $\bar{x}$ means $\Lambda^{-\dim(x)} x$. Using the integral parametric representation for the Heaviside $\Theta$-functions, we get:
\begin{align*}
\nonumber\mathcal{A}_{\mathcal{T}_{n+1}}(p)&=\Lambda^{2\eta}\int_{y_1,y_2}\int_{\vec{q}}\frac{\delta(y_1-\vec{q}\,^{2\eta})}{\vec{q}\,^{2\eta}+\bar{m}^{2\eta}}\,\bigg[\int_{\vec{p}_b} \delta_{p_c\,q_c}\frac{\delta(y_2-\vec{p}_b\,^{2\eta})}{(\vec{p}_b\,^{2\eta}+\bar{m}^{2\eta})^{c(b)+1}}\times \prod_{m=1}^{c(b)}\bar{ \mathcal{A}}_{\mathcal{T}_{p(m)}}(p_{c(m)})\bigg]\\
&=\Lambda^{2\eta}\int_{y_1,y_2}\int_{\vec{q}}\frac{\delta(y_1-\vec{q}\,^{2\eta})}{y_1+\bar{m}^{2\eta}}\,\bigg[\int_{\vec{p}_b} \delta_{p_c\,q_c}\frac{\delta(y_2-\vec{p}_b\,^{2\eta})}{(y_2+\bar{m}^{2\eta})^{c(b)+1}}\times \prod_{m=1}^{c(b)}\bar{ \mathcal{A}}_{\mathcal{T}_{p(m)}}(p_{c(m)})\bigg] \,.
\end{align*}
Finally, rescaling the $q_i$ variables as $q_i\to (y_1-q_c^{2\eta})^{1/2\eta} q_i\,\,\forall i\neq c$,
\begin{align*}
\nonumber\mathcal{A}_{\mathcal{T}_{n+1}}(p)&=\Lambda^{2\eta}\iota(d)\int_{y_1}\frac{y_1}{y_1+\bar{m}^{2\eta}}\,\bigg[\int_{y_2,\vec{p}_b} \left(1-\frac{q_c^{2\eta}}{y_1}\right)\frac{\delta(y_2-\vec{p}_b\,^{2\eta})}{(y_2+\bar{m}^{2\eta})^{c(b)+1}}\times \prod_{m=1}^{c(b)}\bar{ \mathcal{A}}_{\mathcal{T}_{p(m)}}(p_{c(m)})\bigg] \,.
\end{align*}
We have $d-c(b)$ different choices for the color $c$ leaving this expression unchanged; then, we can use the same trick as for the computation of $P_2$ for the two-loop contribution. By summing all the possibles choices, and taking into account the properties of the corresponding delta function, we generate a factor $1/(d-c(b))$; such that the term $q_c^{2\eta}/y_1$ can be discarded in the large $d$ limit. As a result, the amplitude becomes:
\begin{align}
\nonumber\mathcal{A}_{\mathcal{T}_{n+1}}(p)&=\Lambda^{2\eta}\iota(d)\omega^\prime(\bar{m}^{2\eta})\,\bigg[\frac{-1}{m(b)}\frac{\partial}{\partial \bar{m}^{2\eta}}\left(\int_{y_2,\vec{p}_b} \frac{\delta(y_2-\vec{p}_b\,^{2\eta})}{(y_2+\bar{m}^{2\eta})^{m(b)}} \right)\prod_{m=1}^{c(b)}\bar{ \mathcal{A}}_{\mathcal{T}_{p(m)}}(p_{c(m)})\bigg] \,.\label{eqrecurr}
\end{align}
The procedure can be continued from the external leafs to the root vertex. In large $d$, we see that all the faces have length one with a very large probability, such that only the root vertex shares the external momenta. Finally, because $\sum_b m(b)=2n-1$, $(-1)^{\sum_b(m(b)-1)}=(-1)^{n-1}$, which ends the proof of the proposition.
\begin{flushright}
$\square$
\end{flushright}

\begin{definition}
We will denote by $\tau^\star(p)$ the part of the two-point function expanding only in terms of the melonics two points amplitudes, keeping only the relevant ones in the large $d$ limit.
\end{definition}

\noindent
Note that the strategy for non-vacuum two-point diagrams can be done for vacuum diagrams as well, allowing to compute the perturbative contribution of the free energy in the same way as the two-point function $\tau^\star$. The free energy with vanish source is:
\begin{equation}
f(\lambda)=\ln \mathcal{Z}(\lambda)\,,
\end{equation}
where, in contrast to $\tau$, $f$ admits a Feynman expansion with amplitudes labeled with vacuum diagrams. In contrast with two point diagrams considered in section \ref{nloop}, vacuum diagrams in the intermediate field representation have no roots:
\begin{equation}
f(\lambda)=\sum_n (-\lambda)^n\sum_{\mathcal{G}_n}\frac{1}{s(\mathcal{G}_n)}\, \mathcal{A}_{\mathcal{G}_n}\,,
\end{equation}
where $\mathcal{G}_n$ are vacuum Feynman diagrams. Except for the absence of rooted loop-vertex, the proof of proposition \ref{propositionbehavior} may be repeated step by step for vacuum diagrams; therefore, we must have:
\begin{corollary}
Let $\mathcal{G}_n$ be a melonic vacuum diagram with $n$-loops and $\mathcal{T}_n$ the corresponding tree in the intermediate field representation. In the large-$d$ limit ($d\gg n$), the relevant contribution may be decomposed as:
\begin{equation}
\mathcal{A}_{\mathcal{T}_n}= \,v_n(\bar{m}^{2\eta})\,(\iota(d))^{n+1}\,,
\end{equation}
where $v_n(\bar{m}^{2\eta})$ depends only on mass and is explicitly given by:
\begin{equation}
c_n(\bar{m}^{2\eta})=(-1)^{n}\left[\prod_{b\in\mathcal{T}_n}\frac{\omega^{(m(b))}(\bar{m}^{2\eta})}{[(m(b)-1)!]} \right]\,.
\end{equation}
\end{corollary}

\subsection{Formal summation theorems}
In this section, we provide the last two relevant results of this paper, i.e. the resummation theorem, leading to an explicit expression for $\tau^\star(p)$. To make the proof clearer, we divide it into two steps, computing $\tau^\star(0)$ as the first step, from which we will deduce $\tau^\star(p)$ in a second time.

\subsubsection{Resummation for $\tau(0)$.}

The perturbative expansion for $\tau(0)$ may be written as a sum over amplitudes indexed by $1PI$ Feynman diagrams $\mathcal{G}_{i}$, with external vertex of color $i$:
\begin{equation}
{\tau}^\star(0)=\sum_{n=1}^\infty (-\lambda)^n \, \sum_{\mathcal{G}_{i,n}} \,\frac{1}{s(\mathcal{G}_{i,n})} \,\mathcal{A}_{\mathcal{G}_{i,n}}\,,
\end{equation}
Where the symmetry factor ${n!}/{s(\mathcal{G}_{1,n})}$ count the number of independent Wick contractions leading to the same graph $\mathcal{G}_{1,n}$, and where the last sum run over Feynman diagrams with $n$ vertices, and external vertex of color $1$. For the rest of this section, we fix $i=1$. Moreover, we focus on the melonic diagrams only, and the large rank limit is restricted to the melonic diagrams having vertices of different colors and we denote by $\mathfrak{M}_{n,d}$ this set. As recalled in Appendix \ref{App1}, melonic diagrams correspond to trees in the HS representation. Moreover, as in the proposition \ref{propositionbehavior}, the amplitudes $\mathcal{A}_{\mathcal{G}_{1,n}}$ depends only on the coordination vertex numbers of the corresponding tree and may be naturally indexed by tree rather than melon diagram. In an abusive notation, we must have $\mathcal{A}_{\mathcal{G}_{1,n}} \equiv \mathcal{A}_{\mathcal{T}_{1,n}}$. Our final aim is then to rewrite the previous sum over $\mathfrak{M}_{n,d}$ as a sum over rooted trees, with the root of color $1$ and edges of different colors. More precisely, denoting as $\mathfrak{T}_{n,d}$ the corresponding set of trees, we have to find $\tilde{s}(\mathcal{T}_{1,n})$ such that:
\begin{equation}
\sum_{\mathcal{G}_{i,n}\in\mathfrak{M}_{n,d}} \,\frac{1}{s(\mathcal{G}_{i,n})} \,=:\sum_{\mathcal{T}_{1,n}\in\mathfrak{T}_{n,d}}\frac{1}{\tilde{s}(\mathcal{T}_{1,n})} \,.
\end{equation}
To compute $n!/\tilde{s}(\mathcal{T}_{1,n})$, we first remark that this factor must be a product of three distinct contributions. The first contribution $2^n$ arises from the two possible orientations for each original quartic vertices, building the edges of the tree. The second factor is a purely combinatorial number counting the number of color arrangements. More precisely, we must have a factor $(n-1)!$ counting the number of different permutations of the internal (original) vertices, arising from Wick contractions. Another factor arises from the expansion of the exponential itself. Indeed, denoting as $a_i$ for $i$ running from $1$ to $d$ the quartic interaction involved in the action, we must have a combinatorial factor counting the number $\mathcal{N}(n,d)$ of the way to build an arrangement of $n$ quartic vertex of different colors (but including the color $1$) among $(a_1+a_2+\cdots+a_d)^n$. It is not hard to check that this number must be equal to:
\begin{equation}
\mathcal{N}(n,d)=n\frac{(d-1)!}{(d-n)!(n-1)!} (n-1)!\,.
\end{equation}
The first factor counts the $n$ different ways to choose the root vertex of color $1$. The central factor, on the other hand, counts the number of ways to choose $n-1$ different colors among the remaining $d-1$, and finally the last factor $(n-1)!$ counts the different arrangements for a given selection of $n-1$ colors. Taking into account all these contributions, we define $1/\tilde{s}^\prime(\mathcal{T}_{1,n})$ such that:
\begin{equation}
\frac{n!}{\tilde{s}(\mathcal{T}_{1,n})}=:\frac{2^n}{d} n! \frac{d!}{(d-n)!} \frac{1}{\tilde{s}^\prime(\mathcal{T}_{1,n})}\,.
\end{equation}
The interest to extract this factor comes from the explicit expression of the leading order amplitudes in large $d$. The amplitude, in fact, does not depend on the selected set of colors for the $n$ edges, so the sum can be reduced on the set $\mathfrak{T}_{n}$ of planar rooted trees with $n$ vertices:
\begin{equation}
\sum_{\mathcal{T}\in \mathfrak{T}_{n,d}}\frac{1}{\tilde{s}(\mathcal{T})} \mathcal{A}_{\mathcal{T}}= \frac{2^n}{d}\, \frac{d!}{(d-n)!}\sum_{\mathcal{T}\in\mathfrak{T}_{n}}\frac{1}{\tilde{s}^\prime(\mathcal{T})} \mathcal{A}_{\mathcal{T}}\,,
\end{equation}
and the zero-momentum two point function $\tau(0)$ can be written as:
\begin{equation}
{\tau}^\star(0)=\sum_{n=1}^{\infty} \frac{(-2\lambda)^n}{d}\, \frac{d!}{(d-n)!}\sum_{\mathcal{T}\in\mathfrak{T}_{n}}\frac{1}{\tilde{s}^\prime(\mathcal{T})} \mathcal{A}_{\mathcal{T}}\,.
\end{equation}
The remaining factor $1/{\tilde{s}^\prime(\mathcal{T})}$ depends only on the combinatorial structure of trees, but not on the specificities of the model. The same factor has to occur for models with a trivial propagator and a single melonic interaction. For such a model, $\mathcal{A}_{\mathcal{T}}=1$, and the computation have to be done explicitly in the melonic sector using Schwinger-Dyson equation \cite{Samary:2014oya}, \cite{Gurau:2013cbh}-\cite{Bonzom:2011zz}. The result is:
\begin{equation}
\Sigma= \frac{1-\sqrt{1+8\lambda}}{2} = \sum_{n=1}^{\infty} (-2\lambda)^n C_{n-1}\,,
\end{equation}
where $\{C_{n}\}$ denote the Catalan numbers. Therefore, $C_{n-1}$ is precisely the number of planar rooted trees with $n$ loop vertices:
\begin{equation}
C_{n-1}\equiv \sum_{\mathcal{T}\in\mathfrak{T}_{n}}\,,
\end{equation}
ensuring $\tilde{s}^\prime(\mathcal{T})=1$. The Catalan numbers $C_{n}$ are defined as:
\begin{equation}
C_{n}=\frac{1}{n+1} \mathcal{C}_{n}^{2n}\,,
\end{equation}
where $\mathcal{C}_p^n$ denotes the usual binomial coefficients $\mathcal{C}_p^n=n!/p!(n-p)!$. The amplitude $\mathcal{A}_{\mathcal{T}}$ depending only on the coordination numbers of the tree, it could be suitable to convert the sum over trees as a sum over modified coordination numbers $\iota_b:=m(b)-1$, satisfying the hard constraint :
\begin{equation}
\sum_b \, \iota_b=n-1\,.
\end{equation}
Moreover, it is not hard to prove that :
\begin{equation}
\sum_{\underset{\sum_b \, \iota_b=n-1}{i_1,\cdots,i_n}}= \mathcal{C}_{n-1}^{2n-2} \, \Rightarrow \, C_{n-1}=\frac{1}{n}\sum_{\underset{\sum_b \, \iota_b=n-1}{i_1,\cdots,i_n}}\,.
\end{equation}
Then, $C_{n-1}$ being the sum over trees, the previous decomposition is nothing but the desired result, a sum over the trees rewritten as a sum over coordination numbers. With this respect, $\tau(0)$ becomes:
\begin{equation}
{\tau}^\star(0)=-\Lambda^{2\eta}\,\frac{1}{d}\sum_{n=1}^{\infty} \,(2\iota(d)\lambda)^n \frac{d!}{(d-n)!}\,\frac{1}{n}\sum_{\underset{\sum_b \, \iota_b=n-1}{i_1,\cdots,i_n}} \prod_{b=1}^n \frac{(\omega^{\prime})^{(\iota_b)}}{\iota_b!} \label{expression}
\end{equation}
where we took into account the proposition \ref{propositionbehavior}. In the large-$d$ limit, we may use the standard Stirling formula $n!\sim \sqrt{2\pi n}n^ne^{-n}$. Now because
\begin{equation}
(d-n)^{d-n}=e^{d(1-n/d)(\ln(d)+\ln(1-n/d))}=d^{d-n}e^{-n}+\mathcal{O}(n/d)\,,
\end{equation}
we must have:
\begin{equation}
\frac{d!}{(d-n)!} = \frac{d^n e^{-d}}{(d-n)^{d-n} e^{-d+n}}+\mathcal{O}(n/d)=d^n+\mathcal{O}(n/d)\,
\end{equation}
and the previous expression \eqref{expression} becomes, introducing the dimensionless function $\bar{\tau}^\star(0)={\tau}^\star(0)/\Lambda^{2\eta}$:
\begin{equation}
\bar{\tau}^\star(0)=-\frac{1}{d}\sum_{n=1}^{\infty} \,\left(2d\iota(d)\lambda \right)^n \,\frac{1}{n !}\sum_{\underset{\sum_b \, \iota_b=n-1}{i_1,\cdots,i_n}} \frac{(n-1)!}{\prod_b \iota_b!} \, \prod_{b=1}^n (\omega^{\prime})^{(\iota_b)}\,. \label{intermediatesum}
\end{equation}
This expression provides the first important intermediate statement. Indeed, we see that each term of the sum involves increasing the power of $d\iota(d)$. Therefore the existence of an interesting large $d$ limit implies the existence of an appropriate rescaling of the coupling constant, ensuring that each leading order term in $1/d$ receives the same weight. The rescaling can be read directly from the previous expression, and we summarize this result as an intermediate statement:
\begin{lemma}
In the melonic sector, the $d\to \infty$ limit exist for the classical action with rescaled coupling $\lambda\to g/d\iota(d)$:
\begin{equation}
S[T,\bar{T}]=\sum_{\vec{p}} \bar{T}_{\vec{p}} \, \mathcal{K}(\vec{p}\,) \, T_{\vec{p}} +\frac{g}{d\iota(d)} \sum_{i} \sum_{\vec{p}_1,\cdots,\vec{p}_4}\mathcal{V}^{(i)}_{\vec{p}_1,\vec{p}_2,\vec{p}_3,\vec{p}_4}T_{\vec{p}_1}\bar{T}_{\vec{p}_2}T_{\vec{p}_3}\bar{T}_{\vec{p}_4}\,.\label{classicaction2}
\end{equation}
\end{lemma}
Now, we move on to the main statement of this section, the resummation theorem, providing an explicit expression for the large $d$ melonic two-point function. The trick to resum the complicated expression given by \eqref{intermediatesum} use the generalized Leibniz formula:
\begin{equation}
\frac{d^n}{dx^n}(f_1f_2\cdots f_m)=\sum_{\underset{\sum_i \, k_i=n}{k_1,\cdots,k_m}} \frac{n!}{k_1!k_2!\cdots k_m!}\prod_i f_i^{(k_i)}\,,
\end{equation}
such that \eqref{intermediatesum} can be rewritten as:
\begin{equation}
-d\bar{\tau}^\star(0)=\sum_{n=1}^{\infty} \,\left(2g\right)^n \,\frac{1}{n !} \frac{d^{n-1}}{dx^{n-1}} (\omega^{\prime}(\bar{m}^{2\eta}+x))^{n}\bigg\vert_{x=0}\,. \label{intermediatesum2}
\end{equation}
This expression leads to a transcendental equation thanks to the well-known Lagrange inversion theorem, which state that:
\begin{theorem}\textbf{Lagrange inversion theorem.} \label{inv}
Let $f$ be a $\mathcal{C}^{\infty}$ function and $z$ be a function of the variables $x$, $y$ and $f$ as:
\begin{equation}
z=x+yf(z)\,.
\end{equation}
Therefore, for any $\mathcal{C}^{\infty}$ function $h$, we must have:
\begin{equation}
h(z)=h(x)+\sum_{k=1}^{\infty} \frac{y^k}{k!} \frac{d^{k-1}}{dx^{k-1}} ((f(x))^kh^{\prime}(x))\,.
\end{equation}
\end{theorem}
Applying this result where $h$ being the identity function, from equation \eqref{intermediatesum2} we get straightforwardly that $z:=-d\bar{\tau}^\star(0)$ must satisfy the transcendental closed equation:
\begin{equation}
z=\left(2g\right)\,\omega^{\prime}(\bar{m}^{2\eta}+z)\,,\quad \omega^{\prime}(x)=1-x\ln \frac{1+x}{x}\,.
\end{equation}
It is not hard to recover the one and two-loop computations, equations \eqref{masscorr} and \eqref{twolooptau}. Indeed, expanding $z$ in power of $g$ as $z=z^{(1)}+z^{(2)}+\cdots$, we get:
\begin{equation}
z^{(1)}=2g\,\omega^{\prime}(\bar{m}^{2\eta})\,,\qquad z^{(2)}=(2g)^2 \omega^{\prime}(\bar{m}^{2\eta})\omega^{\prime\prime}(\bar{m}^{2\eta})\,,
\end{equation}
which coincide respectively with \eqref{masscorr} and \eqref{twolooptau}. Introducing the \textit{effective mass} $u:=\bar{m}^{2\eta}+z$, the closed equation can be rewritten as:
\begin{equation}
u=\left(\bar{m}^{2\eta}+2g\right)-\left(2g\right) u \ln\frac{1+u}{u}\,,
\end{equation}
or, defining $t^{-1}:=1+u^{-1}$:
\begin{equation}
e^{-\frac{1+\bar{m}^{2\eta}+2g}{2g}}t \ln\left(e^{-\frac{1+\bar{m}^{2\eta}+2g}{2g}}t\right)+e^{-\frac{1+\bar{m}^{2\eta}+2g}{2g}}\left(\frac{e\bar{m}^{2\eta}}{2g}+1\right)=0\,.
\end{equation}
This equation can be formally solved in terms of Lambert functions $W(x)$, defined with the following simple relation:
\begin{equation}
W(x)e^{W(x)}=x\,,
\end{equation}
and we get:
\begin{eqnarray}
t=\exp\left[W\left(\Delta\right)+\frac{1+\bar{m}^{2\eta}+2g}{2g}\right]\,,\quad \Delta:=-\left(\frac{\bar{m}^{2\eta}}{2g}+1\right)e^{-\frac{1+\bar{m}^{2\eta}+2g}{2g}}\,. \label{Delta} \label{asymptotic1}
\end{eqnarray}
Strictly speaking, this formula hold in the very large $d$ limit, i.e for $d\to +\infty$. Indeed, the sum over $n$ being for arbitrary large $n$, we expect that the condition $d\gg n$ must be violated for large $n$. Then, to sum over large $n$, we have to assume the convergence of the series \textit{à priori}, and then discard the $1/d$ contributions as sub-leading order. In other words, the formula \eqref{asymptotic1} must be viewed as an asymptotic formula, to which the exact two-point function must converge in the limit $d\to +\infty$. To summarize, we have then proved the following statement:
\begin{proposition}\label{prokey1}
In the very large $d$ limit, the melonic effective mass for the rescaled quartic melonic model given by \eqref{classicaction2} goes asymptotically toward $u^\star$, given by:
\begin{align}
u^\star&:=\left(\exp\left[-W\left(-\left(\frac{\bar{m}^{2\eta}}{2g}+1\right)e^{-\frac{1+\bar{m}^{2\eta}+2g}{2g}}\right)-\frac{1+\bar{m}^{2\eta}+2g}{2g}\right]-1\right)^{-1}\\
&\,\equiv -\frac{\bar{m}^{2\eta}+2g}{2gW\left(-\left(\frac{\bar{m}^{2\eta}}{2g}+1\right)e^{-\frac{1+\bar{m}^{2\eta}+2g}{2g}}\right)+\bar{m}^{2\eta}+2g}
\end{align}
\end{proposition}
This asymptotic formula has to be completed with some important remarks. The Lambert function $W(x)$ is multivalued in the interval $-1/e\leq x \leq 0$. The first branch, usually called $W_0(x)$ is defined on $\mathbb{R}$ for $x\geq -1/e$, whereas the second branch, $W_{-1}(x)$ is defined on $-1/e\leq x \leq 0$. In both cases, the physical region has to be bounded by the condition $\Delta\geq -1/e$, to ensure the reality of the resumed solution. If we only consider positive definite coupling, to ensure the integrability of the partition function, the choice of the solution must depend on the sign of mass. For $1+\bar{m}^{2\eta}/2g \geq 0$, it is not hard to check that only $W_{-1}$ admits a perturbative expansion around $g\to 0$. Indeed, for $x\to 0^-$, $W_{-1}(x)\approx \ln(-x)$, and $u^\star\to \bar{m}^{2\eta}$, which is the result we have expected. In the opposite, for $1+\bar{m}^{2\eta}/2g < 0$, the solution in agreement with the perturbative expansion is $W_0$. Indeed, for large $x$, $W_{0}(x)\approx \ln(x)$, and, for $g\to 0$:
\begin{align*}
u^\star \to \frac{\vert \bar{m}^{2\eta} \vert }{2g\ln\left(\left\vert\frac{\bar{m}^{2\eta}}{2g}+1\right\vert e^{\left\vert\frac{\bar{m}^{2\eta}}{2g}+1\right\vert-1/2g}\right)-2g\left\vert\frac{\bar{m}^{2\eta}}{2g}+1\right\vert}=-\vert \bar{m}^{2\eta} \vert+\mathcal{O}(g)\,.
\end{align*}
To summarize, in the positive region $1+\bar{m}^{2\eta}/2g \geq 0$, $\Delta\geq -1/e$, the two solutions $W_{0}$ and $W_{-1}$ coexist, but only the second one admits the good limit for $g\to 0$. In the negative region $1+\bar{m}^{2\eta}/2g < 0$, $\Delta\geq -1/e$ however, only the solution $W_0$ exist, and admits the expected limit for vanishing coupling. We have then two branches of solutions and a strong discontinuity along the line $2g+\bar{m}^{2\eta} = 0$. Note that, however, the two solutions are continuous at the point $g=0$, as the previous computation has shown explicitly. We will continue this discussion in the last section in which we will extend our solution to arbitrary momenta.

\subsubsection{Solution for arbitrary momentum}
For $\tau^\star(p)$, from proposition \eqref{propositionbehavior}, the expression \eqref{intermediatesum} must be replaced by:
\begin{equation}
-\frac{d}{1-x}\bar{\tau}^\star(x)=\sum_{n=1}^{\infty} \,\left(2g \right)^n \,\frac{1}{n !}\sum_{\underset{\sum_b \, \iota_b=n-1}{i_1,\cdots,i_n}} \left[ \frac{(n-1)!}{\prod_{b\neq r} \iota_b!} \, \prod_{b\neq r} (\omega^{\prime})^{(\iota_b)}\right]\times \mathcal{A}_r(x)\,, \label{intermediatesum3}
\end{equation}
where we took implicitly into account that $\tau^\star(p)$ depends only on $p^{2\eta}$, and introduce the dimensionless variable $x:=p^{2\eta}/\Lambda^{2\eta}$. Moreover, it is easy to check that:
\begin{equation}
\mathcal{A}_r(x)=\frac{1}{\iota_r!} \frac{\partial^{\iota_r}}{\partial (\bar{m}^{2\eta})^{\iota_r}}\, \frac{\partial}{\partial \bar{m}^{2\eta}}\tilde\omega(\bar{m}^{2\eta},x)\,,
\end{equation}
with:
\begin{equation}
\frac{\partial}{\partial \bar{m}^{2\eta}}\tilde\omega(\bar{m}^{2\eta},x) = \int_0^1 dy\,\frac{y}{y+\frac{\bar{m}^{2\eta}+x}{1-x}}\,.
\end{equation}
As a result, the expansion \eqref{intermediatesum3} can be rewritten in a more suggesting form as:
\begin{equation}
-\frac{d}{1-x}\bar{\tau}^\star(x)=-\frac{1}{d}(1-x)\sum_{n=1}^{\infty} \,\left(2g \right)^n \,\frac{1}{n !}\sum_{\underset{\sum_b \, \iota_b=n-1}{i_1,\cdots,i_n}} \left[ \frac{(n-1)!}{\prod_{b} \iota_b!} \, (\tilde{\omega}^{\prime})^{(\iota_r)}\, \prod_{b\neq r} (\omega^{\prime})^{(\iota_b)}\right]\,.
\end{equation}
where the “prime" designates derivative with respect to $\bar{m}^{2\eta}$. Once again, from the generalized Leibniz formula, each term may be rewritten as a single derivative of order $n-1$ acting on a product of functions:
\begin{equation}
-\frac{d}{1-x}\bar{\tau}^\star(x)=\sum_{n=1}^{\infty} \,\left(2g \right)^n \,\frac{1}{n !}\, \frac{\partial^{n-1}}{\partial(\bar{m}^{2\eta})^{n-1}} \,\left((\omega^{\prime})^{n}\Xi^{\prime} \right)\,,
\end{equation}
where we introduced $\Xi^{\prime}$ defined as $\Xi^{\prime}:=\tilde{\omega}^{\prime}/{\omega}^{\prime}$. Therefore, fixing the arbitrary integration constant such that $\Xi^{\prime}(\bar{m}^{2\eta}+y,x)$ vanish for $y=0$; the Lagrange inversion theorem \ref{inv} must be applied, leading to:
\begin{equation}
-\frac{d}{1-x}\bar{\tau}^\star(x)=\Xi(\bar{m}^{2\eta}+z,x)\,,
\end{equation}
where $z$ must be defined as $z=(2g)\omega^{\prime}(\bar{m}^{2\eta}+z)$, which is nothing but $-d\tau(0)$ given in the last section. Therefore, the full asymptotic function $\bar{\tau}^\star(x)$ is essentially the one loop function, where the bare mass is replaced by the effective mass:
\begin{proposition}\label{propkey2}
In the large $d$ limit, and the melonic sector, the momentum depends two-point function $\tau(p)$ goes toward the asymptotic behavior:
\begin{equation}
\bar{\tau}^\star(x)=-\frac{1-x}{d}\Xi(\bar{m}^{2\eta}-d\tau(0),x)\,, \label{closedfull}
\end{equation}
where the function $\Xi$ is defined as:
\begin{equation}
\Xi(y,x):=\int_0^{y} dt \,\frac{\tilde{\omega}^{\prime}(\bar{m}^{2\eta}+t,x)}{\omega^{\prime}(\bar{m}^{2\eta}+t)}\,.
\end{equation}
\end{proposition}

\subsection{Solving C-S equations in the large $d$ limit}

We now move on two the last topic of this section. What we can learn from the previous formula about the global renormalization group flow? The explicit expression for all the beta functions can be obtained directly from propositions \ref{prokey1} and \ref{propkey2}, merged with corollary \ref{cor1} and equation \eqref{betastep}. However, due to the complicated structure of the previous expression, we keep the deep analysis for another work. To conclude this part we focus on the existence of non-Gaussian fixed points. In section \ref{section2}, we showed that, in the melonic sector, all non-Gaussian fixed points have to verify the strong condition $\gamma=0$, $\gamma$ being the anomalous dimension. From this condition, we can investigate the possibility that the $\ beta$ functions $\beta$ and $\beta_{\bar{m}}$ both vanish when $\gamma=0$. To this end, let us consider the C-S equation for effective mass, \eqref{eqbetam2}, replacing the effective mass $\bar{m}^{2\eta}-d\tau(0)$ by the asymptotic solution for large $d$, $\Lambda^{2\eta} u^\star(g,\bar{m}^{2\eta})$:
\begin{equation}
\left( \Lambda \frac{\partial}{\partial \Lambda}+\beta(g)\,\frac{\partial}{\partial g}+(2\eta+\beta_{\bar{m}})\,\frac{\partial}{\partial \bar{m}^{2\eta}} \right)\Lambda^{2\eta} u^\star(g,\bar{m}^{2\eta})=0\,.
\end{equation}
Then, computing each derivative, we get straightforwardly :
\begin{equation}
2\eta u^\star(g,\bar{m}^{2\eta})+\beta(g) \frac{\partial u^\star}{\partial g}+\beta_{\bar{m}} \frac{\partial u^\star}{\partial \bar{m}^{2\eta}}=0\,.\label{equation1}
\end{equation}
Now, let us consider the relation \eqref{betastep} between $\ beta$ functions. This relation is a consequence of the Ward identities and holds for any dimension. Setting $\gamma=0$, and up to the replacement $\beta(\lambda)\to \beta(g)$ and $\lambda\to g/d\iota(d)$, we get:
\begin{equation}
\frac{1}{d\iota(d)}\beta(g) +\frac{1}{d^2\iota(d)}\frac{2g}{1+\bar{m}^{2\eta}-2g/d}\left(\beta(g)-\frac{g}{1+\bar{m}^{2\eta}}\beta_{\bar{m}}\right)=0\,.\label{betastep2}
\end{equation}
It is not hard to check that $\beta$ must be of order $1$ whereas $\beta_{\bar{m}}$ must be of order $d$. Indeed, in contrast with the mass, the radiative corrections for couplings require fixing one color. The same conclusion may be deduced directly from the previous expression. Note that, due to the mass dimension, the expansion of $\beta_{\bar{m}}$ has to start with $-2\eta \bar{m}^{2\eta}\approx -d\bar{m}^{2\eta}/2$. Setting $d$ arbitrarily large, we get:
\begin{equation}
\beta(g)=\frac{1}{d} \frac{2g^2}{(1+\bar{m}^{2\eta})^2}\beta_{\bar{m}}\,.
\end{equation}
Then, from equation \eqref{equation1}, we deduce straightforwardly:
\begin{equation}
\beta_{\bar{m}}=-\frac{ 2\eta u^\star(g,\bar{m}^{2\eta})}{\dfrac{1}{d} \dfrac{2g^2}{(1+\bar{m}^{2\eta})^2} \dfrac{\partial u^\star}{\partial g}+\dfrac{\partial u^\star}{\partial \bar{m}^{2\eta}}} = \frac{-2\eta}{\partial_{\bar{m}^{2\eta}} \ln(u^{\star})}+\mathcal{O}(1/d)\,.
\end{equation}
Therefore, to get a non-trivial fixed point, we must have $\beta_{\bar{m}}=0$. Investigating this condition requires some algebraic manipulations. Computing the derivative of the logarithm using proposition \ref{prokey1}, we get:
\begin{align}
\nonumber \partial_{\bar{m}^{2\eta}} \ln(u^{\star}) = \frac{\partial_{\bar{m}^{2\eta}} W(\Delta)+\partial_{\bar{m}^{2\eta}}a}{e^{-W(\Delta)-a}-1}\,,
\end{align}
where $a:=(1+\bar{m}^{2\eta}+2g)/2g$. From proposition \ref{prokey1}, we then deduce that:
\begin{align}
\partial_{\bar{m}^{2\eta}} \ln(u^{\star})=u^\star\left( W^\prime(\Delta)\partial_{\bar{m}^{2\eta}}\Delta +1/2g \right) = \frac{\partial_{\bar{m}^{2\eta}}\Delta+\frac{\Delta+e^{W(\Delta)}}{2g}}{\Delta+e^{W(\Delta)}}\,,
\end{align}
where we used the well-known formula for the derivative of the Lambert function:
\begin{equation}
W^\prime(x)=\frac{1}{x+e^{W(x)}}\,.
\end{equation}
Therefore, the expression for $\beta_{\bar{m}}$ becomes:
\begin{equation}
\beta_{\bar{m}}=-\frac{2\eta}{u^\star} \frac{\Delta+e^{W(\Delta)}}{\partial_{\bar{m}^{2\eta}}\Delta+\frac{\Delta+e^{W(\Delta)}}{2g}}\,. \label{equationbatar}
\end{equation}
The derivative of $\Delta$ can be easily computed, leading to:
\begin{equation}
\partial_{\bar{m}^{2\eta}}\Delta=\left[-\frac{1}{2g}+\bigg(\frac{\bar{m}^{2\eta}}{2g}+1\bigg)\frac{1}{2g}\right]e^{-a}=\frac{1}{2g}\bigg(\frac{\bar{m}^{2\eta}}{2g}\bigg)e^{-a}\,,
\end{equation}
and equation \eqref{equationbatar} becomes:
\begin{equation}
\beta_{\bar{m}}=-\frac{4\eta g}{u^\star} \frac{\Delta+e^{W(\Delta)}}{-e^{-a}+e^{W(\Delta)}}=\frac{4\eta g}{u^\star}\frac{\Delta+e^{W(\Delta)}}{e^{-W(\Delta)-a}-1}\,e^{-W(\Delta)}\,.
\end{equation}
Now, from proposition \eqref{prokey1}, the denominator $e^{-W(\Delta)-a}-1$ is nothing but $1/u^\star$. Then, we finally deduce the following corollary:
\begin{corollary}
To any fixed point in the deep UV region ($\Lambda \gg 1$), the melonic beta functions $\beta(g)$ and $\beta_{\bar{m}}$ have to satisfy asymptotically, for very large $d$:
\begin{equation}
\beta_{\bar{m}}=4\eta g(1+\Delta e^{-W(\Delta)})=4\eta g(1+W(\Delta))\,,
\end{equation}
and
\begin{equation}
\beta(g)=\frac{2g^3}{(1+\bar{m}^{2\eta})^2}(1+W(\Delta))\,.
\end{equation}
where $\Delta$ given by equation \eqref{Delta}.
\end{corollary}
We are now in a position to investigate the existence of a non-Gaussian fixed point. From the elementary properties of the Lambert-W function, $\Delta+W(\Delta)$ vanishes only for $\Delta=-1/e$ (see Figure \ref{courbe} below). This point, however, has been pointed out to be the boundary of the analytical region, beyond it the Lambert function takes complex values and the resummation breakdown. Moreover, the condition $\Delta=-1/e$ is a global condition on the boundary and not an isolated point. Therefore:
\begin{claim}
In the large $d$ limit, and the melonic sector, there are no isolated fixed points in the interior of the perturbative region $\Delta > -1/e$.
\end{claim}

\begin{figure}
\begin{center}
\includegraphics[scale=0.7]{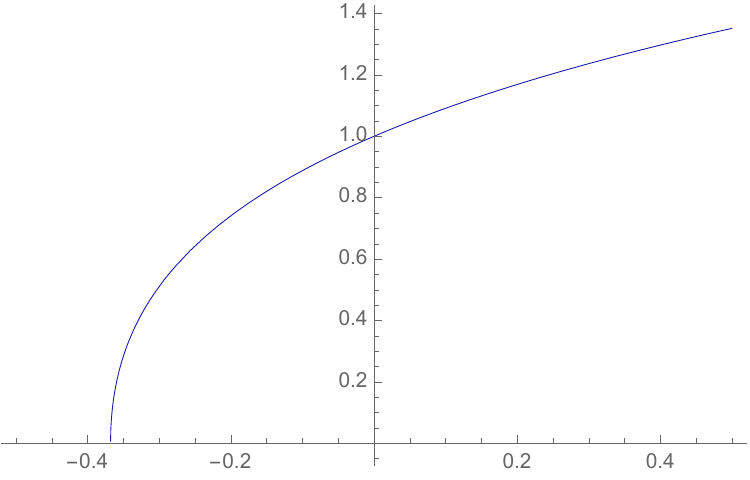}
\captionof{figure}{Numerical plot of the function $f(x)=xe^{-W(x)}+1$. } \label{courbe}
\end{center}
\end{figure}

\section{Discussion and conclusion}
In this paper, we investigated a new family of tensorial group field theories, just renormalizable for arbitrary ranks $d$. From Ward Takahashi identities, we showed that, for any $d$, a strong relationship exists between $\beta$-functions in the deep UV limit, using C-S equation, from which we deduce that any melonic non-Gaussian fixed point must have a vanishing anomalous dimension (the C-S equation is used here because of its flexibility without a choice of a certain regulator, but the same result may be obtained by using the Wetterich flow equation or other RG methods \cite{Berges:2000ew}-\cite{Iliopoulos:1974ur} which is deserved to forthcoming investigation). Similar relations have been deduced recently in the functional renormalization framework, and all tentative to merge this constraint with approximate melonic solutions of the exact renormalization group flow equations lead to the same conclusions: the disappearance of the non-Gaussian melonic fixed point \cite{Lahoche:2018ggd}-\cite{Lahoche:2019orv}. This result remains a claim, due to the necessity to use approximations to solve the renormalization group equations. In all cited papers, the principal approximation is given by the necessity to use the derivative expansion to obtain a tractable parametrization of the theory space. The difficulty, as mentioned in \cite{Lahoche:2018oeo} comes essentially, in the melonic sector and may be translated as the difficulty to solve the closed equation satisfied by the two-point function. In this paper, considering the large $d$ limit, we had able to obtain an asymptotic formula for the bare two-point function, solving the closed equation in the same limit by construction. Then, using the C-S formalism, we deduced the explicit expressions of the $\beta$ functions, to all orders of the perturbation theory at the points where the anomalous dimension vanishes. From an analytic expression, it is clear that in the considered limit, no isolated fixed points occur in the analytic region where the perturbative expansion can be resumed as a real function. \\

\noindent

The results discussed in this introductory paper do not exhaust this novel topic. First of all, we do not investigate the behavior of the full RG flow, we focused only on the regions with vanishing anomalous dimensions. Indeed, the disappearance of isolated fixed points is not the end of history; and a rigorous analysis of the RG behavior has to be done in the future. Related to this point, the method used to obtain the $\ beta$ function, using the Callan-Symanzik equation is crudely rudimentary, and a more sophisticated approach exists to build the RG flow. The solution for the two-point function, for instance, can be used to improve the truncations abundantly used to solve the functional RG equations. Other approaches, using discrete slicing in the momentum space have to be considered for TGFT these last years and could be investigated beyond the one-loop order using the large-$d$ limit. The nature of the limit in itself should be studied carefully. Indeed, our resumed formula provides only the asymptotic behavior of the correlation function in the large rank limit; but we have no control over the neglected contributions of order $1/d$, which can become relevant when $n$, the order of the perturbative expansion, and $d$ are commensurable. This situation is reminiscent of what occurs for large $N$ expansions for matrix and tensor models around the critical point, leading to the double scaling limit investigations. Then, this question, and more generally the existence of a true $1/d$ expansion have to be addressed for incoming work. Finally, the existence of two branches of solution for the resumed two-point function has to be investigated as well. As we will see in section \ref{section3}, the two branches are continuous at the Gaussian fixed point, but a finite gap exists for finite $\vert \bar{m}^{2\eta}\vert $. Moreover, the fact that the two solutions coexist in the region $\Delta\leq 0$ seems two indicate that along a certain curve passing through the origin, we pass continuously from a picture with two vacuum to a picture with one vacuum state, but with a strong discontinuity for other points along the line $g=0$. This qualitative picture is reminiscent of a first-order phase transition, with the Gaussian fixed point playing the role of a critical point. The possible existence of such a transition has been discussed in some recent papers \cite{Lahoche:2018hou}, using approximates solutions for the RG flow. However, at this stage, it is too early to view our result as a definitive statement, which needs to be confirmed by extensive work.

\pagebreak
\begin{appendices}
\section{Power counting and renormalizability}\label{App1}

In this section we provide the power counting for our deformed family of the model, and we will recall some basic properties of the leading order graphs, the so-called \textit{melons}. The proofs are standard, and we will give only the relevant details for the unfamiliar reader. For more details see \cite{Samary:2012bw} and \cite{Geloun:2013saa}.

\subsection{Multi-scale analysis}
We start by fixing our notations. First, we introduce an integer $\rho$ and a positive real number $M$ so that $\Lambda=M^\rho$. Then we define the sharp momentum cutoff $\chi_{\le\rho} (\vec p\,) $, equal to 1 if $ \vec p\,^{2\eta} \le M^{2\eta\rho}$ and zero otherwise which is nothing but the Heaviside step function. The theory with sharp ‘‘cutoff $\rho$" is defined using the covariance
\begin{equation}
C^{\rho}(\vec p) =C(\vec p\,) \chi_{\le\rho} (\vec p\,).
\end{equation}
Then, the key strategy of the multiscale analysis is to slice the theory according to :
\begin{equation}
C^{\rho}(\vec p\,) =\sum_{i=1}^\rho C_i (\vec p\,),\; C_i (\vec p\,) = C(\vec p\,) \chi_i ( \vec p\,^{2\eta}) \label{cutoffsha}
\end{equation}
where $\chi_1$ is 1 if $ \vec p\,^{2\eta} \le M^{2\eta}$ and zero otherwise and for $i\ge 2$ $\chi_i$ is 1 if $M^{2\eta(i-1)} < \vec p\,^{2\eta} \le M^{2\eta\, i}$ and zero otherwise. Now, we need to define the notion of subgraph. A subgraph $S \subset \mathcal G$ in an initial Feynman graph is a certain subset of dotted edges (propagators $C$) with the vertices hooked to them; the half-edges attached to the vertices of $S$ (whether external lines of $G$ or half-internal lines of $G$ which do not belong to $S$) form the external edges of $\mathcal G$.\\

\noindent
Decomposing each propagator into slices, multi-scale decomposition attributes a scale to each line $\ell \in \mathcal{L} (\mathcal {G}) $ of any amplitude $\mathcal{A}_{\mathcal{G}}$ associated to the Feynman graph $\mathcal {G}$. Let us start by establishing multi-scale power counting. \\

\noindent
The amplitude of a graph $ \mathcal {G} $, $ \mathcal {A}_\mathcal {G} $, with fixed external momenta, is thus divided into the sum of all the scale attributions
$ \mu = \{i_{\ell}, \ell \in \cL ( \mathcal {G} ) \} $, where $i_\ell$ is the scale of the momentum $p$ of line $\ell$:
\begin{equation}
\mathcal{A}(\mathcal{G})=\sum_{\mu}\mathcal{A}_{\mu}(\mathcal{G}).
\end{equation}
At fixed scale attribution $\mu$, we can identify the power counting as the powers of $M$. The essential role is played by the
subgraph $\mathcal{G}_i$ built as the subset of dotted edges of $\mathcal{G}$ with scales higher than $i$. From the momentum conservation rule
along any loop vertex, this subgraph is automatically a PI subgraph
which decomposes into $k(i)$ connected PI components: $\mathcal{G}_i=\cup_{k=1}^{k(i)} \mathcal{G}_i^{(k)}$. Note that the inclusion relations between these connected components
indexed by the pair $(i,k)$ build the tree which is called Gallavotti-Nicol\`o tree. We have :
\begin{theorem}
The amplitude $\mathcal{A}_{\mu}(\mathcal{G})$
is bounded by:
\begin{equation}
\quad|\mathcal{A}_{\mu}(\mathcal{G})|\leqslant K^{L({\mathcal{G}})} \prod_{i}\prod_{k=1}^{k(i)}M^{\omega(\mathcal{G}^k_i)} , \ K>0 ,
\end{equation}
and the divergence degree $\omega(\mathcal{H})$
of a connected subgraph $\mathcal{H}$ is given by:
\begin{equation}
\omega(\mathcal{H})=-2\eta L(\mathcal{H})+F(\mathcal{H}),
\end{equation}
where $L(\mathcal{H})$ and $F(\mathcal{H})$ are respectively the number of lines and internal faces of the subgraph $\mathcal{H}$.
\end{theorem}

\noindent
\textit{Proof.} First we have the trivial bounds (for $K=M^{2\eta}$):
\begin{equation}
\vert C_i(\vec{p}\,) \vert \leq K M^{-2\eta i} \chi_{\le i } ( \vec p\,).
\end{equation}
Then, fixing the external momenta for all external faces, the Feynman amplitude (in this momentum representation) is bounded by
\begin{align}
|A_{\mu}(\mathcal{G})|\leq & \left[ \prod_{\ell\in\mathcal{L}(\mathcal{G})}KM^{-2\eta i_\ell}\right]
\prod_{f\in F_{int}(\mathcal{G})} \sum_{p_f\in {\mathbb Z}} \prod_{\ell \in \partial f}
\chi_{\le i_\ell } ( \vec p) \,,
\end{align}
which is deduced straightforwardly from the standard Feynman rules. Then, as a first step, we distribute the powers of $M$ to all the $\mathcal{G}_i^{(k)}$ connected components.
To this end, we note that: $M^i=M^{-1}\prod_{j=0}^i M$, implying:
$\prod_{\ell\in L(G)}M^{-2\eta i_l}=M^{2\eta}\prod_{\ell\in L(G)}\prod_{i=0}^{i_\ell}M^{-2\eta}$. Then, inverting the order of the double product leads to
\begin{equation}
\prod_{\ell\in L(\mathcal{G})}M^{-2\eta i_\ell}=\prod_i\prod_{\ell\in \mathcal{L}(\cup_{k=1}^{k(i)}\mathcal{G}_i^k)}M^{-2\eta}=\prod_i\prod_{k=1}^{k(i)}\prod_{l\in \mathcal{L}(\mathcal{G}_i^k)}M^{-2\eta}=\prod_{i,k}M^{-2\eta L(\mathcal{G}_i^k)}.
\end{equation}
The final step is to optimize the weight of the sum over the momenta $p_f$ of the internal faces.
Summing over $p_f$ with a factor $\chi_{\le i } ( \vec p\,)$ leads to a factor $KM^i$, hence we should sum with the smallest values $i(f)$ of slices $i$ for the lines $\ell \in \partial f$
along the face $f$. This is exactly
the value at which, starting from $i$ large and going down towards $i=0$ the face becomes first internal for some $\mathcal{G}_i^k$.
Hence in this way we could bound the sums $\prod_{f\in F_{int}(\mathcal{G})} \sum_{p_f\in {\mathbb Z}} $ by
\begin{equation}
\prod_i\prod_{k=1}^{k(i)}M^{F(\mathcal{G}_i^k)}.
\end{equation}
Identifying the exponent with $\omega(\mathcal{G}_i^k)$ for each connected component $\mathcal{G}_i^k$, we conclude the proof.
\begin{flushright}
$\square$
\end{flushright}
\subsection{Leading order graphs}

\begin{figure}
\begin{center}
\includegraphics[scale=0.8]{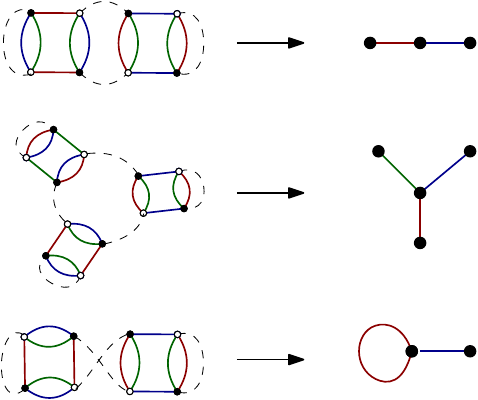}
\captionof{figure}{Correspondence between original representation (on left) and intermediate field representation (on right).}\label{FigApp1}
\end{center}
\end{figure}

To discuss the leading order sector, we introduce an alternative representation of the theory, called \textit{intermediate field representation}, in which the properties of the leading sector become very nice. Usually, intermediate field representation is introduced as a ‘‘trick" coming from the properties of the Gaussian integration and allowing to break a quartic interaction for a single field as a three-body interaction for two fields. To simplify the presentation, we introduce the intermediate field decomposition as a one-to-one correspondence between Feynman graphs \cite{Rivasseau:2017xbk}-\cite{Rivasseau:1991ub} see also \cite{Lionni:2018cnl}-\cite{Rivasseau:2013ova} and references therein. In this section moreover, we only focus our discussion on the vacuum graphs. First, to each vertex of type $i$, we associate an edge of the same color. Second, to each loop made with a cycle of doted edges, we associate a black node, whose number of \textit{corners} corresponds to the length of the loop (Figure \ref{FigApp1} provides some illustrations). To distinguish this representation from the standard Feynman one, we call \textit{colored edges} the edges of the Feynman graphs in the intermediate field representation, and \textit{loop-vertices} their nodes. \\

\noindent
The main statement is then the following:
\begin{theorem}
The 1PI leading order vacuum graphs are trees in the intermediate field representation. Moreover, $4\eta$ must be equal to $d-1$ for a just renormalizable theory. We call melonic diagrams these trees. \label{theoremtreees}
\end{theorem}

\noindent
\textit{Proof.} First of all, consider the case of a 1PI vacuum graph. If it is a tree made with $n$ loop vertices, it must have $c=2(n-1)$ corners, and $F=(d-1)n +1 $ faces since each colored edge glues two faces. As a result, $\omega=-4\eta(n-1)+(d-1)n+1=[d-1-4\eta]n+(1+4\eta)$. \\

\noindent
Then consider a graph with $q$ colored edges, which is not necessarily a tree. For $q=1$ there are two typical configurations:
\begin{equation}
\vcenter{\hbox{\includegraphics[scale=1]{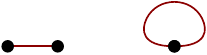} }}
\end{equation}
and so one for each choice of colors for the intermediate field edges. From direct computation, the divergent degrees are respectively, from left to right: $\omega_L=2[d-1-4\eta]+(1+4\eta)$ and $\omega_R=\omega_L-(d-2)$; then the leading order graph is the one on the left, which is a tree. Now, starting with a tree for arbitrary $q$, we have to investigate all the different ways to build a graph with $q+1$ colored edges. From the typical tree
\begin{equation}
\vcenter{\hbox{\includegraphics[scale=1]{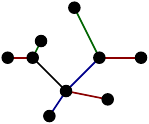} }}\,,
\end{equation}
we have four possible moves:
\begin{equation}
\vcenter{\hbox{\includegraphics[scale=0.8]{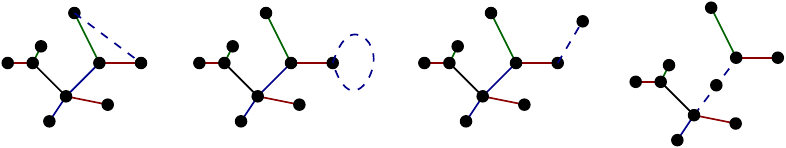} }}
\end{equation}
where the moves are pictured with dotted edges. The two moves on the right preserve the tree structure, and then, the power counting is the expected one for such a tree: $\omega_T=-4\eta(n-1)+(d-1)n+1=[d-1-4\eta](n+1)+(1+4\eta)$. The two moves on the left however both introduce a loop. For the first one, we create at least a single face and two corners. The variation for power counting is then optimally:$\delta\omega=-4\eta+2$. These bounds hold for the second move on the left which creates a tadpole edge. Then for these two moves, we have the bound:
\begin{equation}
\omega \leq [d-1-4\eta](n+1)+(1+4\eta)-(d-1-4\eta)+\delta\omega=\omega_T-(d-3)\,.
\end{equation}
As a result, the power counting is bounded by trees for $d>3$. Finally, if need to have a just-renormalizable leading sector, the divergent degrees do not increase with the number of loop vertices. We then require $d-1-4\eta=0$, implying $\eta>1/2$.
\begin{flushright}
$\square$
\end{flushright}

\noindent
The leading order non-vacuum graphs can be obtained following a recursive procedure. To this end, we have to keep in mind the definitions \ref{def1} of Section \ref{section23}, which we complete with the following:
\begin{definition}
The heart graph of a melonic 1PI Feynman graph $\mathcal{G}$ is the subset of vertices and lines obtained from the deletion of the external vertices.
\end{definition}
\noindent
Now, consider a vacuum melonic diagram. We obtain a two-point graph by cutting one of the dotted edges. Due to the structure of melonic diagrams, it is clear that if we cut an edge which is not a tadpole (i.e. an edge in a loop of length upper than one), we obtain a 1PR diagram. Cutting a first tadpole edge, we delete $d$ faces. $d-1$ of them become boundary external faces while the other one becomes a heart external face. We then obtained a 1PI two points melonic diagram. Then to obtain a four points melonic diagram, we have to cut another tadpole edge on this diagram. However, it is clear that such a cutting deletes $d$ internal faces, \textit{except} if the chosen tadpole is on the path of the opened heart external face. Indeed, in this case, the cutting $d-1$ faces (which become boundary external faces) for the same in dotted lines, and the power counting is optimal. Moreover, due to the deletion, we created another heart external face, obviously of the same color as the one for the original two-point diagram. Recursively, we deduce the following proposition:
\noindent
\begin{proposition} \label{cormelons}
A 1PI melonic diagram with $2N$ external lines has $N(d-1)$ external faces of length $1$ shared by external vertices and $N$ heart external faces of the same color running through the internal vertices and/or internal lines (i.e. through the heart graph).
\end{proposition}
To complete these definitions, and of interest for our incoming results, we have the following proposition:
\begin{proposition}
All the divergences are contained in the melonic sector.
\end{proposition}
Pointing out that all the counter-terms in the perturbative renormalization are fixed from the melonic diagrams only. Proof may be found in \cite{BenGeloun:2011rc}. Finally, we can add an important remark about melonic diagrams: Their divergent degrees depend only on the number of external edges, as expected for a just-renormalizable theory. To be more precise, note that the number of dotted edges is related to the number of vertices as $2L=4V-N_{ext}$, where $N_{ext}$ denotes the number of external dotted edges. Moreover, it is easy to see, from the recursive definition of melons that $F=(d-1)(L-V+1)$. Indeed, starting from a Feynman graph in the original representation, it is obvious that contracting a tree (dotted) edge does not change the number of faces. Then, contracting all the edges over such a spanning tree, we get $L-V+1$ remaining edges, hooked on a single vertex, building a \textit{rosette}. Now, we delete the edges optimally, following successive $(d-1)$-dipole contractions. We recall that a $k$-dipole is made with two black and white nodes (in the original representation), wished together with one dotted edge and $k$ colored edges. In the intermediate field representation, we can then start from a vacuum diagram, and proceed both with the dipole and tree contractions. Starting from a leaf, hooked to an effective vertex $b$ with $p$ external edges hooked to him, we can contract the leaf, discarding $(d-1)$ faces and $1$ dotted edge. We may assume that only leafs are hooked to $b$, except for one colored edge. Using the same procedure for all the leafs, we get an effective loop of length $p$, on which we can contract $p-1$ edges to get a new tadpole, that we can contract, and so on. Repeating the same procedure for all loop-vertex, we get $F=(d-1)(L-V+1)+1$. For a non-vacuum graph with $2N$ external edges, creating them cost $d-1$ faces per deleted tadpole, except for the first one, which cost $d$ faces, and the desired result follows. From this counting for faces, the divergence degree becomes
\begin{align}\label{melocountinplus}
\nonumber\omega=-2\eta L+F&=-2\eta(2V-N_{ext}/2)+(d-1)(V-N_{ext}/2+1)\\
&=\big[(d-1)-4\eta\big]V+\bigg[(d-1)-\bigg(\frac{d-1}{2}-\eta\bigg)N_{ext}\bigg]\,.
\end{align}
which is nothing but the relation \eqref{dimrel}.

\section{The key sums with sharp regulator }\label{App2}
In this section, we derive the important sums that arise in the computation of the loop expansion of the two-point correlation function. Consider the following sum:
\begin{equation}
S_1(p=0,a,b)=\sum_{\vec{q}\in \mathbb{Z}^{d-1}}\frac{\Theta(\Lambda^{2\eta}-\vert\vec{q}\,^{2\eta}\vert)}{a\vert\vec{q}\,^{2\eta}\vert+b}\,.
\end{equation}
In the large $\Lambda$ limit and by introducing the continuous variable $x=q/\Lambda$ we get the following integral representation
\begin{align}
\nonumber S_1(p=0,a,b)\approx I_1(p=0,a,b)&=2^{d-1}\Lambda^{2\eta}\int_{\mathbb{R}^{+d-1}} d^{d-1}x\frac{\Theta(1-\vec{x}^{2\eta})}{a\vec{x}\,^{2\eta}+b'}\\
&=2^{d-1}\Lambda^{2\eta}\int_{0}^1dy\int_{\mathbb{R}^{+d-1}} d^{d-1}x\frac{\delta(y-\vec{x}^{2\eta})}{a\vec{x}\,^{2\eta}+b'}\,,
\end{align}
with $b'=b/\Lambda^{2\eta}$. Using the properties of the delta distribution, we find:
\bea
I_1(p=0,a,b)&=&2^{d-1}\Lambda^{2\eta}\left(\int_{0}^1\frac{ydy}{ay+b'}\right)\int_{\mathbb{R}^{+d-1}} d^{d-1}x\delta(1-\vec{x}^{2\eta})\cr
&=&\Lambda^{2\eta}\iota(d) \frac{1}{a}\bigg[1-\frac{b'}{a}\ln\left(\frac{a+b'}{b'}\right)\bigg]
\eea\label{Stepinter}
with:
\begin{equation}
\iota(d):=2^{d-1}\int_{\mathbb{R}^{*d-1}} d^{d-1}x\delta(1-\vec{x}^{2\eta})=2^{d-1}\int_0^1dx_1\cdots \int_0^1 dx_{d-1}\delta(1-\vec{x}^{2\eta})\,.
\end{equation}
This integral can be computed using Feynman parameters formula ($\Re(\alpha)>0$):
\begin{equation}
\frac{1}{A_1^\alpha\cdots A_{d-1}^\alpha}=\frac{\Gamma((d-1)\alpha)}{[\Gamma(\alpha)]^{d-1}}\int_0^1 du_1\cdots du_{d-1} \delta\left(1-\sum_i u_i\right)\frac{\prod_i u_i^{\alpha-1}}{(\sum_{i=1}^{d-1} A_i u_i)^{(d-1)}}\,, \label{Feynman}
\end{equation}
with $A_i=1\,\forall i$:
\begin{equation}
\iota(d)=2^{d-1}\bigg[\Gamma\left(\frac{d+1}{d-1}\right)\bigg]^{d-1}\,.
\end{equation}\label{iota}
Then:
\begin{equation}
I_1(p=0,a,b)=2^{d-1}\Lambda^{2\eta} \bigg[\Gamma\left(\frac{d+1}{d-1}\right)\bigg]^{d-1}\frac{1}{a}\bigg[1-\frac{b'}{a}\ln\left(\frac{a+b'}{b'}\right)\bigg] \,.\label{formulaK}
\end{equation}
In the same way, we define:
\begin{equation}
S_2(p=0,a,b)=\sum_{\vec{q}\in \mathbb{Z}^{d-1}}\frac{\Theta(\Lambda^{2\eta}-\vert\vec{q}^{2\eta}\vert)}{[a\vert\vec{q}\,^{2\eta}\vert+b]^2}\,,
\end{equation}
and
\begin{equation}
S_2(p=0,a,b)=-\frac{d}{db}S_1(p=0,a,b)\,,
\end{equation}
providing the integral approximation:
\begin{equation}
I_2(p=0,a,b)=2^{d-1}\bigg[\Gamma\left(\frac{d+1}{d-1}\right)\bigg]^{d-1}\frac{1}{a^2}\bigg[\ln\left(\frac{a+b'}{b'}\right)-\frac{a}{a+b'}\bigg]\,.\label{formulaL}
\end{equation}

\end{appendices}

\end{document}